\documentclass[11pt]{article}

\usepackage{amssymb,latexsym, amsmath,amscd,graphicx}
\usepackage[all,cmtip]{xy}

\makeatletter
\@addtoreset{figure}{section}
\def\thefigure{\thesection.\@arabic\c@figure} \def\fps@figure{h, t}
\@addtoreset{table}{bsection}
\def\thetable{\thesection.\@arabic\c@table} \def\fps@table{h, t}
\@addtoreset{equation}{section}

\makeatother

\newcommand{\R}{\mathbb{ R}}
\DeclareMathOperator{\tr}{tr}

\newtheorem{thm}{Theorem}[section]
\newtheorem{prop}[thm]{Proposition}
\newtheorem{lem}[thm]{Lemma}

\newtheorem{rem}{Remark}

\def\la{\lambda}
\DeclareMathOperator{\Prym}{Prym}
\DeclareMathOperator{\Jac}{Jac}
\DeclareMathOperator{\diag}{diag}
\DeclareMathOperator{\Ad}{Ad}

\begin{document}
\title{Continuous and discrete Neumann systems on Stiefel varieties as
matrix generalizations of the Jacobi--Mumford systems
}
\author{Yuri N. Fedorov \and Bo\v zidar Jovanovi\'c}

\maketitle

\begin{abstract}
We study geometric and algebraic geometric properties of
the continuous and discrete Neumann systems on cotangent bundles of Stiefel varieties $V_{n,r}$.
The systems are integrable in the non-commutative sense, and by applying a $2r\times 2r$--Lax representation,
we show that generic complex invariant manifolds are open subsets of affine (non-compact) Prym varieties on which
the complex flow is linear. The characteristics of the varieties and the direction of the flow are calculated
explicitly.
Next, we construct a family of (multi-valued) integrable discretizations of the Neumann systems and describe them
as translations on the Prym varieties, which are written explicitly in terms of divisors of points on the spectral curve.
It appears that the systems inherit or naturally generalize the basic properties of the classical Neumann system
on $S^{n-1}$ and, therefore, of the Jacobi--Mumford systems: the structure of the Lax matrices,
the spectral curve, the equations of motion, linearization on Abelian varieties, and, in the discrete case,
the formula for the translation on them.
\end{abstract}

\tableofcontents

\section{Introduction}
The classical integrable Neumann system on an $n$-dimensional sphere
is known to be a toy model for the Jacobi--Mumford systems introduced in \cite{Mum} and describing
translationally invariant flows on the Jacobians of hyperelliptic curves and represented by $2\times 2$ Lax pairs.

During the last two decades various generalizations of the
Jacobi--Mumford systems associated with families of $r$-gonal and
generic curves were constructed (see \cite{Prev, Shil, Ped_Vanh,
Jap_2007}). The paper \cite{FVh} also introduced such systems
linearized on special types of Prym varieties isomorphic to
Jacobians. All such generalizations are closely related to
integrable flows on coadjoint orbits of the loop algebra
$\widetilde{\rm gl}(r)$, with Lax matrices being rank $r$
perturbations of a constant matrix. The complete algebraic
geometric description of such flows in the generic case was given
in the series of papers \cite{Dub_0, AHH0, AHH1, AHH2}.

There have also been several important results on discretizations
(B\"acklund transformations) of the above systems, which are
described by translations on the Jacobians of hyperelliptic,
$r$-gonal, and generic curves, and which can be related to
addition theorems for meromorphic functions on such Jacobians (see
\cite{Ves, KV, Fe1}).

The present paper contributes to these fields of interest in two ways.

First, as a further generalization of the Jacobi--Mumford systems,
we consider natural analogs of the Neumann system on the Stiefel
variety $V_{n,r} $ ($r<n$), the set of $n \times r$ matrices $X$
satisfying the constraints $X^TX= {\bf I}_r$. These analogs were
first introduced by Reiman and Semenov by presenting a ''big''
($n\times n$) Lax pair \cite{RS}. It appears that for $r>1$
the big Lax pair itself does not define yet the system uniquely:
there exists a family of noncommutatively integrable Neumann flows with different
$SO(n)$-invariant metrics $ds^2_\kappa$ on $V_{n,r}$ that possess
the same big Lax representation and share the same foliation on invariant isotropic tori of the cotangent bundle $T^*V_{n,r}$ (see \cite{FJ2008}).

The Neumann systems on $V_{n,r}$ also admit ''small'' ($2r \times
2r$) Lax representations  which, modulo the action of a discrete
group of reflections $\mathbb Z^n_2$, define these systems uniquely (see
\cite{Kap89, FJ2008}). They have the symplectic block form
\begin{gather*}
\dot L(\lambda)
=[L (\lambda), N_\kappa(\lambda)],  \quad
L (\lambda)=\begin{pmatrix}
X^T( \lambda \mathbf{I}_n-A)^{-1} P & X^T(\lambda \mathbf{I}_n-A)^{-1} X \\
\mathbf I_r - P^T(\lambda \mathbf{I}_n- A)^{-1} P & - P^T(\lambda \mathbf{I}_n-A)^{-1} X
\end{pmatrix} ,
\end{gather*}
where $A=\diag (a_1,\dots,a_n)$, $P$ is the $n \times r$-momentum
satisfying the constraint $X^T P + P^T X=0$, $\la$ is a rational
spectral parameter, and the matrix $N_\kappa(\lambda)$, which is not shown
here, depends on the metric $ds^2_\kappa$ (see Section 3).

The matrix $L(\la)$ can be regarded as a direct generalization of
the $2\times 2$ Lax matrix for the classical Neumann system (see
\cite{Mum, AMV, Jap_2007}) and fits to the already studied class
of rank $d$ perturbations of the constant matrix $A$. The latter
are described by the Lax matrices of the form
$$
Y + \sum_{i=1}^n \frac {{\cal N}_i}{\la-a_i}, \qquad  Y, {\cal
N}_i \in \mathrm{gl}(d),
$$
where $Y$ is constant and ${\cal N}_i$ depend on the variables of
the corresponding dynamical system. However our $L(\la)$ belongs
to the symplectic loop subalgebra $\widetilde{\rm
sp}(2r)\subset\widetilde{\rm gl}(2r)$ and has a quite specific
structure, moreover, $Y$ is not diagonalizable, it contains a
Jordan block. As a result, the spectral curve $\cal S$ of $L(\la)$
has extra strong singularities at the infinity (see Section 4).
So, the previous results of \cite{AHH0, AHH1, AHH2} concerning the
genus of $\cal S$, number of its infinite points, etc, need to be
modified in order to treat this case.

Calculating the order of singularity and the genus of the
regularized curve $\cal S$ (Theorem \ref{big_g}), taking into account the
involution $\sigma\, : {\cal S} \longrightarrow {\cal S}$, we then
demonstrate that, up to the action of $\mathbb Z^n_2$, generic
complex invariant manifolds ${\cal I}_h$ of the Neumann systems on
$T^*V_{n,r}$ are open subsets of {\it affine} Prym varieties
$\widetilde{\Prym} ({\mathcal S}, \sigma)$ of the same dimension
 (Theorem \ref{MAIN}). The latter are algebraic noncompact
subgroups of generalized Jacobian varieties $\widetilde{\Jac}
({\mathcal S},\infty)$, the extensions of $\Jac ({\mathcal S})$ by
$\left( {\mathbb C}^*\right)^{[r/2]}$. We show that the
trajectories of the Neumann systems are straight lines on
$\widetilde{\Prym} ({\mathcal S}, \sigma)$ and calculate their
direction (Theorem \ref{linear2}).

It should be noted that affine Prym varieties appear as complex invariant manifolds in several classical integrable systems
on Lie groups {\it before their reduction}, for example,
the spatial rigid body motion in the Clebsch case of the Kirchoff equations (see e.g., \cite{Bel}). However,
to our knowledge, in the existing literature such manifolds were not given the corresponding algebraic description.

Next, the Marsden--Weinstein reductions of the Neumann systems on
$T^*V_{n,r}$ by the group $SO(r)$ are also integrable, and their
complex invariant tori (again, modulo $\mathbb Z^n_2$--action),
are shown to be usual Prym varieties ${\Prym}({\mathcal S},
\sigma)\subset \Jac({\cal S})$ (Proposition \ref{sub_Prym}).  At the level of
the Prym varieties, the reduction gets a particularly transparent
form: it corresponds to the projection $\widetilde{\Prym}
({\mathcal S}, \sigma)\to {\Prym}({\mathcal S}, \sigma)$. In particular,
the reduction for the zero value of the momentum mapping is the Neumann system on
the Grassmann variety $G_{n,r}$, which is integrable in the
usual commutative (Liouville) sense (see Theorem \ref{LinGr}).

Second, we construct a family of (multi-valued) integrable
discretizations (B\"ack\-lund transformations) ${\mathfrak B}_r \,
:\, (X,P)\to (\tilde X, \tilde P)$ of the Neumann system on $T^*
V_{n,r}$ (Section 6). The family is parameterized by $\la_*\in
{\mathbb C}$ and described by the $2r\times 2r$ intertwining
relation (discrete Lax pair)
$$
\tilde L(\la) M (\la, \la_*)= M (\la, \la_*) L(\la), \quad
M (\la, \la_*) = \begin{pmatrix} - \Gamma (\la_*) & {\bf I}_r \\
 (\la - \la_*) {\bf I}_r +  \Gamma^2 (\la_*) &  - \Gamma (\la_*) \end{pmatrix},
$$
where $\tilde L(\la)$ depends on $\tilde X, \tilde P$ in the same
way as $L(\la)$ depends on $X, P$, whereas $\Gamma(\la_*)$ depends
on $X, \tilde X$ in a symmetric way (Theorem \ref{disr_Lax_Neum}).  Again, the
above relation is a direct matrix generalization of the $2\times
2$ discrete Lax pair describing a B\"acklund transformation of the
classical Neumann system (see \cite{HKR, KV}).

Each discretization map ${\mathfrak B}_r$ preserves the first
integrals of the continuous system and geometrically is described
by translations $\cal T$ on $\Prym ({\mathcal S}, \sigma)$
(Theorems \ref{NDI}, \ref{shift}). The translations are written explicitly in
terms of $r$ non-involutive points on the spectral curve $\cal S$
over the coordinate $\la_*$ (i.e., eigenvalues of $L(\la_*)$). A
choice of such points (partition) fixes the branch of the map and
the corresponding translation $\mathcal T$. Iterations of all
possible translations generate a lattice of rank $2^{r-1}$ on the
universal covering of $\Prym ({\mathcal S}, \sigma)$. The above
construction is similar to that used in \cite{MV} to describe the
integrable discrete Frahm--Manakov top on $SO(n)$ and is actually
based on the procedure of solving a matrix quadratic equation
presented earlier in \cite{Pot}.


In Conclusion we discuss further possible generalizations.
Long technical proofs of some theorems are given in Appendix.
For the completeness of the exposition in Section 2 we recall on some basic definitions concerning
Jacobi--Mumford systems and the discretisation of the Neumann problem on the sphere, while in Section 3
we recall on the integrability of continuous Neumann systems studied in \cite{FJ2008}.

\section{The Jacobi--Mumford and the Neumann systems}

\subsection{The standard (odd order) Jacobi--Mumford systems}
Let $\Gamma=\Gamma(R)$ be a smooth hyperelliptic curve of genus
$g$, whose affine part is given by the equation $\mu^2=R(\la)$,
where $R(\la)$ is a {\it monic} polynomial of degree $2g+1$. We
regard $\Gamma$ as a compact Riemann surface having one infinite
point $\infty$. Consider a generic divisor of $g$ points
$P_1=(\la_1,\mu_1), \dots, P_g=(\la_g,\mu_g)$ on $\Gamma$.
Following Mumford \cite{Mum}, the curve and the divisor can
be associated to three polynomials $U(\lambda), V(\lambda)$, and
$W(\lambda)$, $\lambda\in {\mathbb C}$ such that
\begin{eqnarray}
\label{cond1}
&& U(\lambda_i)=0, \qquad V(\lambda_i)=\mu_i, \qquad i=1,\dots,g, \\
&& W(\la)U(\la)+V^2(\la)=R(\la). \label{match}
\end{eqnarray}

The degrees of $U(\lambda)$ and $V(\lambda)$ are {\it at least}
$g$ and $g-1$ respectively. For a fixed curve $\Gamma$ given by
$R(\la)$, the polynomial $W(\la)$ is uniquely determined from the condition (\ref{match}).
Then, if $U(\la)$ is monic of degree $g$, then $W(\la)$ is monic and has degree $g+1$. That is,
\begin{align}
& U(\la)=(\la-\la_1)\cdots(\la-\la_g)=\la^g+u_1\la^{g-1}+\cdots+u_g, \notag \\
& V(\la)=\sum_{k=1}^{g}\mu_k \prod_{j\ne k} \frac{\la-\la_k}{\la_k-\la_j}=v_1\la^{g-1}+\cdots+v_{g}, \label{odd}\\
& W(\la)=\la^{g+1}+ w_0\la^{g} + w_1\la^{g-1} + \cdots+w_g=
(\la-\nu_1)\cdots (\la-\nu_{g+1}). \notag
\end{align}

Now let ${\cal E}_g$ be a variety of all non-constant coefficients of $U(\la),V(\la),W(\la)$.
As shown in \cite{Mum}, for a fixed $R(\la)$,
formula (\ref{match}) defines a set of
equations on the coefficients which provides a purely algebraic
description of an affine part of Jac($\Gamma$), the Jacobian variety of the curve $\Gamma$.
The coefficients themselves are meromorphic functions on the Jacobian.

The variety ${\cal E}_g$ itself can be completed to the fiber
bundle $ \bar{\cal E}_g   \longrightarrow \mathcal R $ over the
$2g+1$-dimensional space $\mathcal R$ spanned by the coefficients
of $R(\lambda)$ and parameterizing odd order hyperelliptic curves
$\Gamma=\Gamma(R)$ of genus $g$ (the base), with fibers of
$\bar{\cal E}_g$ being the Jacobians of the curves.

Any translationally invariant vector flow on Jac$(\Gamma)$ can be
extended (in different ways) to a vector flow on the whole space
$\bar{\cal E}_g$, which leaves the fibers invariant. The latter
flow is described as an integrable system of differential
equations on the coefficients of the three polynomials, which is
referred to as a Jacobi--Mumford or an (odd order) {\it master} system
(see e.g., \cite{Mum, Vanh1, Gavr}).

Let $R\in\mathcal R$, $\Gamma=\Gamma(R)$ be the associated
hyperelliptic curve and let $ {\cal A} \, : \Gamma \longrightarrow
\Jac (\Gamma)$ be the Abel mapping given by the following integral with the basepoint $\infty$
$$
{\cal A} (P) = \int_\infty^P (\omega_1,\dots,\omega_g)^T, \quad
P\in \Gamma,
$$
where $\omega_1,\dots,\omega_g$ are  independent holomorphic
differentials on the curve $\Gamma$.

Consider a translationally invariant flow on Jac$(\Gamma)$ which
is tangent to the image of the curve ${\mathcal
A}(\Gamma)\subset\Jac(\Gamma)$ at
 a finite point $\mathcal A(\la_*,\mu_*)$.
Then, fixing $\la_*\in {\mathbb C}$, but not $R\in\mathcal R$, we
get a vector flow on ${\cal E}_g$ or on $\bar{\cal E}_g$.

The flow can be represented in the following Lax form with $\la$
as a rational parameter (see \cite{Vanh1}):
\begin{align}
\label{LaxP} \dot {\bf L}(\la)=&[{\bf L}(\la), N(\la)]  \\
{\bf L}(\la) =& \begin{pmatrix} V(\la) & U(\la)  \\
   W(\la) & -V(\la)  \end{pmatrix}, \nonumber \\
N(\la) =& \frac 1{2(\la-\la^*)}
                 \begin{pmatrix} -V(\la^*) & -U(\la^*) \\
  -W(\la^*)-(\la-\la^*) U(\la^*) &  V(\la^*) \end{pmatrix}\nonumber\\
=&-\frac{{\bf L}(\la^*)}{2(\la-\la^*)} -\frac 12 \begin{pmatrix}
0 & 0 \\ -U(\la^*) & 0 \end{pmatrix}. \nonumber
\end{align}

Concerning the Lax equation \eqref{LaxP}, the coefficients of the
spectral curve
$$
\det({\bf L}(\la)-\mu {\bf
I}_2)=\mu^2-V^2(\la)-W(\la)U(\la)=\mu^2-R(\lambda)=0
$$
are the first integrals of the system, and we recover the family
of hyperelliptic curves $\Gamma=\Gamma(R)$. When $\la_*$ tends to
$\infty$, after an appropriate time rescaling, the limit flow is
described by \eqref{LaxP} with
\begin{equation}
\label{infinity}
N(\la)= \begin{pmatrix} 0 & 1 \\
                 \la+(w_0-u_1) & 0 \end{pmatrix}.
\end{equation}

\begin{rem}{\rm
To a fixed curve $\Gamma$ and the given divisor $P_1+ \dots +P_g$
on it, one can associate the {\it dual divisor} $R_1+\dots+
R_{g+1}$ given by the zeros of the meromorphic function $
f(\la,\mu)={(\mu+V(\la))}/{U(\la)}. $ In view of (\ref{cond1}) and
(\ref{odd}), $f(\la,\mu)$ have simple poles at the original
divisor and $\infty$, which implies
\begin{equation}
\label{dual} {\cal A}(P_1)+\cdots+{\cal A}(P_g)+{\cal A}(\infty)=
{\cal A}(R_1)+\cdots+{\cal A}(R_g)+{\cal A}(R_{g+1}) .
\end{equation}
As follows from (\ref{match}), one can also write
$f=W(\la)/(\mu-V(\la))$. Hence, the coordinates $\nu_k=\la(R_k)$,
$k=1,\dots, g+1$ are the roots of $W(\la)$. }\end{rem}

\paragraph{The Neumann system on $S^{n-1}$.} The best known example of an
integrable problem associated to the Jacobi--Mumford system is the
Neumann system describing the motion of a point on the unit sphere
$S^{n-1}=\{\langle q,q\rangle=1\}\subset {\mathbb R}^n$,
 with the quadratic potential $V(q)=
-\frac12\langle q, A q\rangle $, $A={\rm diag}(a_1,\dots,a_n)$
(see \cite{Neum, Moser}). Upon introducing momentum
$p=\dot q$ and imposing the constraint $\langle
p,q\rangle =0$, the motion is described by the equations
\begin{equation}
\label{Neumann} \dot q=p, \quad \dot p= A q + \nu q,
\quad \nu=-\langle p,p\rangle-\langle q, A q\rangle, \qquad
i=1,\dots,n,
\end{equation}
which possesses the $2\times 2$ Lax representation
\begin{gather}
\label{standard-neumann} \dot L(\la)=[L(\la), N(\la)], \qquad
L(\la)= \begin{pmatrix}  \sum_{i=1}^{n}\frac{q_i p_i}{\la-a_i}  &
\sum_{i=1}^{n}\frac{q_i^2}{\la-a_i} \\
1-\sum_{i=1}^{n}\frac{p_i^2}{\la-a_i}   &
-\sum_{i=1}^{n}\frac{q_i p_i}{\la-a_i} \end{pmatrix}, \\
\qquad\quad N(\la)=\begin{pmatrix}  0 & 1 \\
             \la+\nu(p,q) &  0 \end{pmatrix} . \label{comp}
\end{gather}

Let ${a}(\la)=(\la-a_1)\cdots (\la-a_n)$. For the polynomial Lax
matrix
${\bf L}(\la)=a(\la) L (\la)$
 the characteristic equation $|{\bf L} (\la) -\mu {\bf I}_2|=0$ has the form
\begin{equation}
\label{U-curve}
\mu^2= - a^2(\la)\left( \sum_{i <j }^n
\frac{(q_i p_j-q_jp_i)^2}{(\la-a_i)(\la-a_j)}
- \sum_{i=1}^n \frac{q_j^2 }{\la-a_i} \right),
\end{equation}
and gives the odd order hyperelliptic curve $\Gamma$ of genus $g=n-1$
\begin{equation}
\label{N-curve} \mu^2= R(\lambda)= {a} (\la)\, (\la-c_1)\cdots
(\la-c_{n-1}).
\end{equation}

For real $(q,p)$, the zeros $c_1,\dots,c_{n-1}$ of $R(\lambda)$ are real (see  Lemma
4.5 in Chapter III of \cite{Mum}) and they represent commuting
integrals of the Neumann system.

 One can then identify the $2\times 2$ polynomial matrices
${\bf L}(\la)=a(\la) L (\la)$ and (\ref{LaxP}) by setting $g=n-1$, which
gives
\begin{align*}
& U(\la)  = {a} (\la ) \sum_{i=1}^{n}\frac{q_i^2}{\la-a_i} = \la^{g}+ (-\tr A+\langle q,A q \rangle) \la^{g-1} + \cdots, \\
& V(\la)  = {a}(\la ) \sum_{i=1}^{n}\frac{q_i p_i}{\la-a_i} = \langle p,Aq\rangle \la^{g-1} + \cdots, \\
& W(\la)  = {a} (\la )
\left(1-\sum_{i=1}^{n}\frac{p_i^2}{\la-a_i}\right) = \la^{g+1} -
(\tr A+\langle p, p\rangle) \la^{g} + \cdots,
\end{align*}
which, in particular, implies
\begin{equation} \label{u_w}
u_1 = -\tr A + \langle q,Aq \rangle ,  \quad v_1=\langle
p,Aq\rangle, \quad w_0 = - \tr A- \langle p,p\rangle.
\end{equation}
Then the second matrix \eqref{comp} coincides with
\eqref{infinity}. Hence the Neumann flow is linearized on the
Jacobian of $\Gamma$, on which it is tangent to $\mathcal
A(\Gamma)\subset \Jac(\Gamma)$ at its infinite point.

The above relations also give the parametrization of $q_i^2$ in
terms of $\la_1,\dots,\la_g$, the $\la$-coordinates of the points
$P_1,\dots, P_g$ on $\Gamma$, which now play the role of the
spheroconical coordinates on $S^{n-1}$:
\begin{equation*}
\label{spheroconic}
q_i^2=\frac{(a_i-\la_1)\cdots (a_i-\la_{n-1})}{\prod_{j\ne i}(a_i-a_j)},
\qquad i=1,\dots,n
\end{equation*}
In addition, $p_i^2$ can be expressed in terms of
$\la$-coordinates of the dual divisor $R_1,\dots,R_{g+1}$:
$$
p_i^2=\frac{(a_i-\nu_1)\cdots (a_i-\nu_{n})}{\prod_{j\ne i}(a_i-a_j)},
\qquad i=1,\dots,n.
$$

Since $p_i^2,\,q_i^2,\,p_i q_i$ are linear functions of the
coefficients of $U,V,W$, they are meromorphic function on
Jac$(\Gamma)$. However, the coordinates $p_i, q_i$ themselves do
not have this property. According to \cite{Mum}, the following
theorem holds

\begin{thm} \label{tori_r=2}
\begin{description}
\item{1)} Complex invariant manifolds of the Neumann system with
the constants of motion $c_i$ factorized by the action of the
discrete group ${\mathbb Z}_2^n$ generated by reflections
$(q_i,p_i)\,\to\, (-q_i,-p_i)$, $i=1,\dots,n$ are open subsets of
the Jacobian $\Jac (\Gamma)$ of the spectral curve
(\ref{N-curve}).

\item{2)} The complex invariant manifolds themselves are open
subsets of unramified coverings of $\Jac(\Gamma)$ obtained by
doubling some of the period vectors of $\Jac (\Gamma)$.
\end{description}
\end{thm}

\subsection{Discretizations of the Neumann system on $T^*S^{n-1}$}

The first integrable discretization of the Neumann system
(\ref{Neumann}) was found in \cite{Ves} by using the approach of
Lagrange correspondences and the idea of factorization of Lax
operators (see \cite{MV} for the details).
Combining the results of \cite{KV, Su}, below we consider the
discretization map $\mathfrak B\colon T^*S^{n-1}(q,p) \to T^*S^{n-1}({\tilde q},{\tilde p})$ containing an extra parameter
$\la_*\in{\mathbb C}$
written in the implicit form\footnote{The above notation slightly differs from that of
Chap. 21.6 of \cite{Su}, where one should replace $h^2$ and
$\Omega$ by  $1/\la^*$ and $-A$, respectively. }
\begin{gather} \label{moms}
p = A^{1/2}(\la_*) \tilde q - \gamma q, \quad \tilde p = - A^{1/2}(\la_*) q +  \gamma\tilde q,
\end{gather}
or, in the form
\begin{gather} \label{sem2}
\tilde q = A^{-1/2}(\la_*) (\gamma  q +p) , \quad \tilde p = -
A^{1/2}(\la_*) q + A^{-1/2}(\la_*) (\gamma^2 q +\gamma  p),
\end{gather}
where $ A(\la_*)= \la_*
{\bf I}_n-A=\diag(\la_*-a_1,\dots,\la_*-a_n)$, and $\gamma$ is a multiplier.

The equations \eqref{sem2} were obtained in
\cite{HKR} in the context of B\"ackund transformations.
As was shown in \cite{HKR}, up to the action of the group of
reflections $(p_i,q_i)\to (-p_i,-q_i)$, the above map is
equivalent to the following intertwining relation (discrete Lax
pair):
\begin{equation}
\label{Kuz}
{\tilde L}(\la) M(\la|\la_*)=M(\la|\la_*)L(\la),
\end{equation}
where $L(\la), {\tilde L}(\la)$ have the same structure as
in (\ref{standard-neumann}) and depend on the ``old'' variables $(p,q)$
and the ``new'' ones $({\tilde q},{\tilde p})$ respectively, and
\begin{equation}
\label{MKuz}
M(\la | \la_*)=\begin{pmatrix} -\gamma  & 1 \\
    \lambda-\la_* +\gamma ^2 &  -\gamma  \end{pmatrix}.
\end{equation}
Indeed, setting in \eqref{Kuz} subsequently
$\lambda=a_1,\dots,a_n$ and calculating the matrix $
 M(\la|\la_*) \, {\mathbf L}(\la)\, M^{-1}(\la|\la_*) \vert_{\la=a_i},
$ one obtains
\begin{equation} \label{pq_scalar}
\tilde q_i = \pm \, \frac {\gamma  q_i+p_i}{\sqrt{\la_* - a_i}},
\quad \tilde p_i = \pm \, \frac {(\gamma^2 + a_i-\la_*) q_i+\gamma
p_i}{\sqrt{\la_* - a_i}} , \qquad i=1,\dots,n ,
\end{equation}
which is equivalent to (\ref{sem2}) if we fix sign "+" above.

To find the multiplier $\gamma$ as a function of $(q,p)$, one applies the condition
$\langle \tilde q, \tilde q\rangle =1$ to the first equation
of \eqref{sem2} and obtains the quadratic equation
\begin{equation} \label{quad_beta}
\langle q, A^{-1}(\la_*) q \rangle \, \gamma^2 + 2 \langle p, A^{-1}(\la_*) q \rangle \, \gamma +
\langle p, A^{-1}(\la_*) p \rangle -1=0 ,
\end{equation}
which gives
$$
\gamma = \frac {- \langle p, A^{-1}(\la_*) q \rangle \pm
\sqrt{\langle p, A^{-1}(\la_*) q \rangle^2- \langle q,
A^{-1}(\la_*) q \rangle \, (\langle p, A^{-1}(\la_*) p
\rangle -1) } }{ \langle q, A^{-1}(\la_*) q \rangle }\, ,
$$
or, in view of (\ref{U-curve}), \eqref{N-curve},
\begin{equation} \label{beta_2}
\gamma = \frac {- \langle p, A^{-1}(\la_*) q \rangle +
\mu(\la_*)/{a} (\la_*)}{ \langle q, A^{-1}(\la_*) q \rangle } =
\frac{-V(\la_*) + \mu(\la_*) }{U(\la_*)} ,
\end{equation}
$\mu(\la_*)$ being the coordinate of one of the two points on the
curve \eqref{N-curve} over $\la=\la_*$.

As a result, the map ${\mathfrak B}$ preserves the same first integrals as the continuous
Neumann system for any $\lambda^*$, and, in view of Theorem \ref{tori_r=2}, its
complex invariant varieties are open subsets of unramified
coverings of the Jacobian of the curve $\Gamma$ given by
\eqref{N-curve}. As also follows from (\ref{beta_2}), the map
${\mathfrak B}$, as well as its inverse ${\mathfrak B}^{-1}$,
is generally 2-valued.

\begin{rem}\label{REAL}{\rm Obviously, equations
(\ref{sem2}) describe a {\it real} map when  $\mu(\la_*)$ is real,
i.e., $R(\la_*)>0$. In particular, $R(\la_*)>0$ when
$\lambda_*>a_1,\dots,a_n,c_1,\dots,c_{n-1}$ (see \eqref{N-curve}).
 }\end{rem}

\begin{rem}\label{shift-rem}{\rm Using the above formulas, one can prove that
the composition of the two branches of ${\mathfrak B}$, corresponding to different choices of sign of $\mu(\la_*)$,
gives the map $(q,p) \to (-q,-p)$. In Section \ref{DDis}
an analog of this property for the discrete system on $V_{n,r}$ will be considered.}
\end{rem}

Algebraic geometrical interpretation of the map (\ref{sem2}),
was given in \cite{Ves} (see also \cite{MV}) and, in
the context of discretization of the Jacobi-Mumford systems, in
\cite{KV}. Namely, let $\bar {\mathfrak B}$ be the reduction of
${\mathfrak B}$ under the action of the reflection group $\mathbb
Z_2^n$. Hence, following Theorem \ref{tori_r=2}, its generic
invariant varieties are open subsets of $\Jac(\Gamma)$. Then, according to the above remark, the two
branches of $\bar{\mathfrak B}$ just represent a ''shift back'' and a
''shift forward'' maps.

Let, as above, ${\cal A} \, : \Gamma \longrightarrow \Jac
(\Gamma)$ be the Abel map.  In \cite{Ves} and \cite{KV} the
following theorem was proven.

\begin{prop} \label{shift0} The restriction of $\bar {\mathfrak B}$ onto $\Jac(\Gamma)$ is
described by translation by the vector
\begin{equation}
\label{shift2} T= {\mathcal A} ({P}) -{\mathcal A} (\infty) =
{\mathcal A} ({P}) , \qquad {P}=(\la_*,\mu_*).
\end{equation}
\end{prop}

Notice that, although the translation vector does not depend
explicitly on the constants of motion, it depends on them via the
moduli of the curve $\Gamma$. The above formula allows
to write explicit solution for the trajectory $\{(q_{k},p_{k}), k\in \mathbb Z\}$, in terms
of hyperelliptic theta-functions associated to $\Gamma$, whose
arguments depend linearly on the discrete time $k \in {\mathbb
Z}$. As follows from Proposition \ref{shift0}, the two
branches of $\bar {\mathfrak B}$ corresponding to
$P=(\la_*,\mu_*), (\la_*, -\mu_*)$ are just shifts by opposite
vectors $\pm T$. This observation is consistent with Remark
\ref{shift-rem} above.

\paragraph{Special cases $\la_* =a_j$.} According to formula (\ref{shift2}), when
$(\la_*,\mu_*) =(a_j,0)$ the translation $\mathcal T$ is just a
half-period in $\Jac(\Gamma)$ and, in view of (\ref{beta_2}), the
map ${\mathfrak B}$ is single-valued. As expected,
double iteration $\bar{\mathfrak B}^2$ gives the same point in
the Jacobian. On the other hand,
$( \tilde{\tilde q}, \tilde{\tilde p})= {\mathfrak B}^2(q,p)=(-q,-p)$.

This can be checked directly: let us set $\la_*=a_j$ in
(\ref{pq_scalar}). Since the denominator in $j$-th equations
vanishes, for $\tilde q_j$ to be finite, one has to set $\gamma
=-p_j/q_j$. Then the second equation in (\ref{pq_scalar}) gives
$\tilde p_j=\gamma \tilde q_j$, i.e., $\tilde p_j/\tilde
q_j=\gamma$, which, under the next iteration of ${\mathfrak B}$,
implies $\tilde\gamma=-\gamma$. Using this relation and iterating
(\ref{sem2}) directly we get $\tilde{\tilde q}=-q$, $\tilde{\tilde
p}=-p$ for any $p,q$. In this sense, the original map ${\mathfrak B}$ with $\la_*
=a_j$ represents the ''imaginary unit map''.

\paragraph{Continuous Limit.} As also follows from (\ref{shift2}), the vector $T$ tends to zero when $\la_*\to \infty$, which must give the continuous limit of the map (\ref{sem2}).
We shall describe this limit in details in Section \ref{DDis}, jointly with the discrete Neumann system on the Stiefel
variety $V_{n,r} $.

\section{The Neumann systems on the Stiefel variety
$V_{n,r}$}\label{NSV}

\paragraph{Definitions.}
The Stiefel variety $V_{n,r}$ is the variety of ordered sets of
$r$ orthogonal unit  vectors $e_1,\dots, e_r$ in the Euclidean
space ${\mathbb R}^n$,  or,
the set of $n \times r$ matrices $X=(e_1 \cdots
e_r)$ satisfying the constraints $X^TX= {\bf I}_r$. It is a smooth $(rn-r(r+1)/2)$--dimensional submanifold in the space
of $n\times r$ real matrices $M_{n,r}=\R^{nr}$. Also, since
the left $SO(n)$--action on $V_{n,r}$ is transitive, it is a homogeneous space
$SO(n)/SO(n-r)$.

The cotangent bundle $T^*V_{n,r} $ can be realized as the set of
pairs of $n\times r$ matrices $({X},{P})$, $P=(p_1 \cdots p_r)$,
that satisfy the constraints
\begin{equation} \label{cond_XP}
{X}^T {X}={\bf I}_r, \quad {X}^T {P} + {P}^T {X} =0\, .
\end{equation}
The canonical symplectic structure $\omega$ on $T^*V_{n,r}$ is the
restriction of the 2-form $\omega_0=\sum_{i=1}^n
\sum_{s=1}^r d{p}_{s}^i\wedge \, d{e}_{s}^i$ from the ambient space
$T^*M_{n,r}$. It is convenient to work with
the redundant variables $(X,P)$ and the corresponding
Dirac--Poisson structure $\{\cdot,\cdot\}$ (see \cite{Dirac, Moser}) on $T^*V_{n,r}$ is
described in \cite{FJ2008}.

The Lie groups $SO(n)$ and $SO(r)$ naturally act on $T^*V(n,r)$ by
left and right multiplications respectively, with the equivariant
momentum mappings given by
\begin{eqnarray*}
&&\Phi: T^*V(n,r) \to \mathrm{so}(n), \,\, \Phi=P X^T-X P^T\,(=p_1\wedge e_1+\dots+p_r\wedge e_r), \label{momentum_map} \\
&&\Psi: T^*V(n,r) \to \mathrm{so}(r), \,\, \Psi=X^T P - P^T X.
\label{momentum_right_map}
\end{eqnarray*}

Note that the actions of $SO(n)$ and $SO(r)$ commute and, in
particular, the components of the momentum mapping, $\Phi_{ij}$
and $\Psi_{ij}=\langle e_i,p_j\rangle-\langle e_j,p_i\rangle$
are $SO(r)$ and $SO(n)$--invariant functions, respectively.
Here we identified $\mathrm{so}(n)\cong \mathrm{so}(n)^*$ and
$\mathrm{so}(r)\cong \mathrm{so}(r)^*$ by the use of the invariant
metrics on $\mathrm{so}(n)$ and $\mathrm{so}(r)$ defined by
$\langle \eta_1,\eta_2\rangle=-\frac12 \tr(\eta_1 \eta_2)$.

\paragraph{The Neumann systems.}
By analogy with the system \eqref{Neumann} on the sphere $S^{n-1}$,
one defines {\it a Neumann on the Stiefel variety} $V_{n,r} $ as a
natural mechanical system with an $SO(n)$--invariant kinetic
energy and the quadratic potential function
\begin{equation*}
V=-\frac12\tr(X^TAX)=-\frac12\sum_{i=1}^r \langle e_i,Ae_i\rangle,
\qquad A=\diag (a_1,\dots,a_n). \label{Neumann_potential}
\end{equation*}
In this paper it is assumed that $a_i \ne a_j$, $i\ne j$.

Following \cite{FJ2008},  we consider a family $SO(n)\times
SO(r)$--invariant metrics $ds^2_\kappa$ defined by the kinetic
energy
$$
T_\kappa(X,P)= \frac12\langle \Phi,\Phi\rangle +\frac12 \kappa
\langle\Psi,\Psi\rangle =
\frac12\tr(P^TP)-\left(\frac12+\kappa\right)\tr((X^T P)^2),
$$
$\kappa$ being a parameter ($\kappa>-1$).  In the class of the
metrics $ds^2_\kappa$ we have the {\it normal metric} induced from
a bi-invariant metric on $SO(n)$ ($\kappa=0$) and the {\it
Euclidean metric} induced from the Euclidean metric of the ambient
space $M_{n,r}$ ($\kappa=-1/2$).

The corresponding Hamiltonian flows are given by
\begin{equation}
\begin{aligned}
& \dot X=P-(1+2\kappa) X P^T X, \\
& \dot P=AX+(1+2\kappa) PX^T P+X\Lambda,
\end{aligned}\label{No}
\end{equation}
where the Lagrange multiplier matrix does not depend on $\kappa$:
\begin{equation}
\Lambda=-X^TAX-P^TP. \label{multiplier}
\end{equation}

Note that, due to the $SO(r)$--symmetry, the
momentum mapping $\Psi$ is the integral of the system \eqref{No}.
The difference between the two vector fields in \eqref{No} with
different $\kappa$ is proportional to $X' = -X \, P^T X$, $P'=P\,
P^T X$. Since $X^T P= - P^T X = \frac 12 \Psi$, the latter
describes permanent (steady state) rotations of the vectors
$e_1,\dots, e_r$ in ${\mathbb R}^r=\text{span}(e_1,\dots, e_r)$,
similarly for the vectors $p_1, \dots, p_r$.

The geometry of Riemannian spaces $(V_{n,r},ds^2_\kappa)$ is
studied in \cite{Je}. It appears that the class of the metrics
$ds^2_\kappa$ is suitable for studying other natural mechanical
problems on $V_{n,r}$ (see \cite{FJ2012}).

\paragraph{Lax representations.}
For all $\kappa$, the equations \eqref{No} admit the "big" $n\times n$--Lax
representation with a spectral parameter $\lambda$ (see \cite{FJ2008})
\begin{equation}
\frac{d}{dt} \mathcal L (\lambda)=[\mathcal N (\lambda),\mathcal
L(\lambda)] \label{LA1}
\end{equation}
 where $\mathcal L (\lambda)$
and $\mathcal N(\lambda)$ are given by
$\mathcal L(\lambda)=\lambda\Phi+XX^T+\lambda^2 A$, $\mathcal
N (\lambda)=\Phi+\lambda A$.

On the other hand, up to the action of the discrete group ${\mathbb Z}_2^n$
generated by $n$ reflections
\begin{equation}\label{reflection}
 ( e_1^{i} ,\dots,e_r^{i}, p_1^{i},\dots, p_r^{i}) \longrightarrow
(- e_1^{i} ,\dots,-e_r^{i}, -p_1^{i},\dots, -p_r^{i}), \qquad
i=1,\dots,n
\end{equation}
the Neumann flows \eqref{No} are equivalent to the following "small""
$2r\times 2r$ matrix Lax representations with a rational spectral
parameter $\lambda$
\begin{align}
& \frac{d}{dt} L(\lambda)
=[L (\lambda), N_\kappa(\lambda)], \label{LA3} \\
& L (\lambda)=\begin{pmatrix}
X^T( \lambda \mathbf{I}_n-A)^{-1} P & X^T(\lambda \mathbf{I}_n-A)^{-1} X \\
\mathbf I_r - P^T(\lambda \mathbf{I}_n- A)^{-1} P & - P^T(\lambda
\mathbf{I}_n-A)^{-1} X
\end{pmatrix} , \label{Lax_r}\\
&  N_\kappa(\lambda)=\begin{pmatrix}
(1+2\kappa)\,X^TP & \mathbf{I}_r \\
\lambda \mathbf{I}_r + \Lambda & - (1+2\kappa) P^T X
\end{pmatrix},
\label{LA4}
\end{align}
the matrix $r\times r$ factor $\Lambda$ being already defined in
\eqref{multiplier}.

The Lax representation \eqref{LA1} is closely related
to that for the integrable Clebsch--Perelomov rigid body system
\cite{Pe} and for $r=1$ it was given by Moser in \cite{Moser}. It
belongs to the class of the Lax matrix representations related to
symmetric pairs decompositions of Lie algebras \cite{RS}.
The small Lax pair (\ref{LA3})--(\ref{LA4}) is a direct
generalization of the $2\times 2$ Lax pair
(\ref{standard-neumann}), \eqref{comp}, and it was first given in
unpublished manuscript \cite{Kap89}.

\paragraph{Non-commutative integrability.}
From the Lax matrix $\mathcal L(\lambda)$ we obtain the set of
commuting integrals $\{f_{k,i}(X,P)\}$ defined by
\begin{equation}
\tr(\lambda(PX^T-XP^T)+XX^T+\lambda^2 A)^k=\sum
f_{k,i}\lambda^i, \quad k=1,\dots,n. \label{perelomov}
\end{equation}

Apart from these integrals, which are $SO(r)$--invariant, the
Neumann flows also possess the non-commutative algebra of
integrals $\Psi_{ij}$. As was shown in \cite{FJ2008}, the above systems
are integrable in the non-commutative sense \cite{N, MF2}.
To describe the geometric structure and the dimension of the
invariant tori, consider the Poisson reduced space $T^*V_{n,r}/SO(r)$.

The integrals $\tr(\Psi^{2k})$ induce
Casimir functions $J_k$, $k=1,\dots,[r/2]$  and
a generic symplectic
leaf ${\cal U}_c =\{ J_k=c_k \}\subset T^*V_{n,r}/SO(r)$ has the dimension
\begin{equation}
2l=2r(n-r)+\frac{r(r-1)}2-\left[\frac{r}{2}\right] \, .
\label{dimension}
\end{equation}

In \cite{FJ2008} we
proved that if all the eigenvalues of $A$ are distinct, then \eqref{perelomov} is a complete
commutative set on $\mathcal U_c$.
 As a
result, there are $l+r(r-1)/2$ independent functions within the algebra of integrals $\{f_{k,l}, \psi_{ij}\}$
and its center has
$l+[r/2]$ independent functions. Note that
the invariants $\tr(\Psi^{2k})$ functionally depend on $\{f_{k,l}\}$, so
the center is generated by the integrals \eqref{perelomov}.

Since $ \text{dim } T^*V_{n,r}
=(l+r(r-1)/2)+(l+[r/2])$, by the theorem of the non-commutative
integrability (see e.g., \cite{MF2, N}),
the Neumann flows \eqref{No} are completely integrable in the
non-commutative sense.  The generic motions of the
systems, with the momentum $\Psi$ of the maximal rank, are
quasi-periodic over the isotropic tori of dimension
\begin{equation}
\label{dim_fol} \delta =
l+\left[\frac{r}{2}\right]=\frac12\big(2r(n-r)+\frac{r(r-1)}2-\left[\frac{r}{2}\right]\big)
+\left[\frac{r}{2}\right] .
\end{equation}

There is also an alternative path to the reduced system.
Namely, the symplectic leaves of the Poisson reduced space
$T^*V_{n,r} /SO(r)$ are the Marsden--Weinstein reduced spaces
$\Psi^{-1}(h)/SO(r)_h$, where $ SO(r)_h=\{Q\in
SO(r)\,\vert\,\Ad_Q(h)=h\} $ is the isotropy subgroup of $h\in
\mathrm{so}(r)$. For a regular $h$,
$SO(r)_h$ is the maximal torus $\mathbb T^{[r/2]}$. The
factorization of $\Psi^{-1}(h)$ by $SO(r)_h \cong\mathbb
T^{[r/2]}$ coincides with the corresponding leaf ${\cal U}_c\subset
T^*V_{n,r}/SO(r)$ with $J_k=c_k$, $c_k=\tr(h^{2k})$, $k=1,\dots,
[r/2]$.

In this way we obtain the reduction of the restricted Neumann
flows on $\Psi^{-1}(h)$ to the symplectic manifolds $\mathcal
U_c$. The systems on $\mathcal U_c$ are integrable in the usual
commutative sense: the invariant isotropic tori ${\mathbb
T}^{\delta}$ laying on $\Psi^{-1}(h)$ reduce to the
$l$--dimensional Lagrangian tori ${\mathbb T}^l$ laying on ${\cal
U}_c$:
\begin{equation*} \label{diagr1}
\begin{CD}
 T^*V (r,n) @ < \qquad \Psi= h \qquad <<  \Psi^{-1}(h)  \supset  {\mathbb T}^{\delta}  \\
       @  V  /SO(r)  VV  @  V   VV /{\mathbb T}^{[r/2]}   \\
T^*V (r,n)/SO(r) \quad @ < J_1
=c_1,\,\dots, \, J_{[r/2]}=c_{[r/2]} << {\cal U}_c \supset
{\mathbb T}^{l}
\end{CD}
\end{equation*}


\paragraph{The case $\Psi=0$ and reduction to the Grassmannian variety.}
The {\it oriented Grassmannian} $G_{n,r}$ is the variety of
$r$-dimensional oriented planes passing through the origin in
$\R^n$. It is a quotient space of the Stiefel manifold $V_{n,r}$
with respect to the right $SO(r)$--action via submersion
$\pi(e_1\,e_2\cdots e_r)=e_1\wedge e_2\wedge \cdots \wedge e_r$,
and its cotangent bundle $T^*G_{n,r}$ is symplectomorphic to the
reduced space $\mathcal U_0=\Psi^{-1}(0)/SO(r)$.
Note that $(X,P)$ belongs to
$\Psi^{-1}(0)$ if and only if $X^T P=0$, and all reduced systems have the same kinetic energy given by a {\it
normal metric} on $G_{n,r}$.
In \cite{FJ2008} we proved that the integrals $\{f_{k,i}(X,P)\}$ induce a complete
commutative set on $\mathcal U_0$. Therefore,
the reduced Neumann flow on $T^*G_{n,r}$ is completely integrable
in the usual Liouville sense.

\section{The spectral curve}

The Lax matrix (\ref{Lax_r}) is a particular case of so called rank $\rho$ perturbations
of the constant matrix $A=\diag (a_1,\dots,a_n)$,
which generate Lax pairs of a great variety of integrable systems and have the form
\begin{equation} \label{Nr*}
L(\la) = Y+ \sum_{i=1}^n \frac {{\cal N}_i}{\la-a_i}, \qquad  Y,
{\cal N}_i \in \mathrm{gl}(\rho),
\end{equation}
where $Y$ is constant and ${\mathcal N}_i$ depend on the variables of
the corresponding dynamical system. In our case $\rho=2r$, $\mathcal N_i$ have rank 1,
\begin{equation*}
\mathcal N_i = \big(e_1^i \, \cdots \, e_r^i \, -p_1^i \,\cdots\,  -p_r^i \big)^T
\big(p_1^i \, \cdots \, p_r^i \, e_1^i \, \cdots \, e_r^i \big),
\end{equation*}
and
\begin{equation} \label{Nr}
Y=\begin{pmatrix} 0 & 0 \\ {\bf I}_r & 0 \end{pmatrix}.
\end{equation}

General cases of rank $\rho$ perturbations corresponding to $Y$
with a simple spectrum and the properties of the corresponding
spectral curves were studied in detail in \cite{Dub_0, Prev, AHH1,
AHH2}. The matrix (\ref{Lax_r}) however has a quite special
structure, and its spectral curve $\cal S$ has extra strong
singularities. So, the previous results concerning its genus,
number of infinite points, etc, should be adopted to our case. For
this reason below we give our proper self-contained description.

Namely, multiplying $L (\lambda)$ in (\ref{Lax_r}) by
${a}(\lambda) =(\lambda-a_1)\cdots (\lambda-a_n)$, we obtain the
polynomial Lax matrix
\begin{equation}\label{polynomial-L}
\mathbf{L}(\la)=a(\lambda)L(\lambda),
\end{equation}
defining the spectral curve
\begin{equation} \label{curve}
\mathcal S\subset\mathbb C^2\{\lambda,w\}:  \quad F(\la,w) =
\det(\mathbf{L}(\lambda)-w{\bf I}_{2r})=0.
\end{equation}

By expanding $F(\la,w)$ in $w$ we get
\begin{equation*}
F(\la,w) =w^{2r} + w^{2r-2} {a}(\lambda) {\cal I}_{2}(\lambda)+
\cdots +w^2 {a}^{2r-3}(\lambda) {\cal I}_{2r-2}(\lambda)+
{a}^{2r-1} (\lambda) {\cal I}_{2r}(\lambda),
\end{equation*}
where ${\cal I}_{2l}(\lambda)$ is a polynomial of degree $n-l$ in
$\lambda$ with the leading coefficient $C_l^r=\frac {r!}{l!
(r-l)!}$. Their explicit expressions are given in
\cite{FJ2008}.\footnote{One should replace $\lambda$ and $A$ by
$-\lambda$ and $-A$, respectively, to relate the small Lax matrix
$L(\lambda)$ and integrals $\mathcal I_{2l}(\lambda)$ used here
with the matrix $\mathcal L^*_{neum}(\lambda)$ and integrals
$\mathcal I_{2l}(\lambda)$ used in \cite{FJ2008}.}

Due to the symplectic block structure of ${L}(\lambda)$, the
coefficients at odd powers of $w$ in $F(\la,w)$ are zero, so the
spectral curve has the involution
\begin{equation}\label{sigma}
\sigma \colon  \,(\la,w) \to (\la, -w).
\end{equation}

Note that, although the Lax matrix $L(\lambda)$ is not invariant
under the right $SO(r)$-action, the spectral curve and therefore
all the integrals $\mathcal I_{2l}(\lambda)$ are invariant.
Moreover, by the Wienstein--Aronszjn formula (see \cite{AHH2, Moser}),
the spectral curve \eqref{curve} is birationally equivalent to the spectral curve of the Lax
matrix $\mathcal L$, and the integrals $\{{\cal I}_{2l}(\lambda)\}$ are linear combination of the integrals \eqref{perelomov}.

\subsection{Genus of the regularized spectral curve.}

From now on we consider the spectral curve ${\cal S}$ in \eqref{curve} as its projective closure in
${\mathbb P}^2=(\xi : \eta : \zeta)$ such that $\la=\xi/\zeta$, $w=\eta/\zeta$.
First, note that in its finite part, $\cal S$ has singular points
$$
{S}_i=(\la =a_i, w=0), \qquad i=1,\dots,n,
$$
where all the branches of
$w(\la)$ meet and all the partial derivative of $F(\la,w)$ vanish
up to order $2r-2$.

Let ${\cal S}'$ be a complete regularization of $\cal S$.
To regularize $\cal S$ at $S_i$, we observe that the eigenvectors of the polynomial matrix
$\mathbf L(\lambda)$ for $\la=a_i$ are proportional to those of ${\cal N}_i$ in (\ref{Nr*}). As was shown in \cite{AHH2}, if $\mathrm{rank}({\cal N}_i)=k_i$, then $\mathbf{L}(a_i)$ has $2r-k_i$ independent
eigenvectors. In our case any ${\cal N}_i$ has rank 1 and $2r-1$ independent eigenvectors. Hence, each
$S_i \in {\cal S}$ gives rise to $2r-1$ different points on ${\cal S}'$, one of them being ordinary (second order) branch point ${P}_{a_i}$ of the covering ${\cal S}' \to {\mathbb P} \{\lambda\}$.

Following the theory of singularities (see e.g., \cite{Kir}), this
implies that at ${S}_i$, the curve $\cal S$ has singularity
($\delta$-invariant) of order
$$
\delta_i = (2r-1)(r-1), \qquad i=1,\dots,n.
$$

\paragraph{Singularity at the infinity.}
The curve ${\cal S}\subset {\mathbb P}^2=(\xi : \eta :
\zeta)$ has only one infinite point $(0:1:0)$. Indeed, the
structure of $F(\la,w)$ in \eqref{curve} implies that $w \cong
O\left( \la^{(2n-1)/2} \right)$ as $\la \to \infty$.
Then, in the neighborhood of $(0:1:0)$,
it is convenient to use local coordinates $t,{\mathfrak w}$ such that
\begin{equation} \label{tW}
 \la =\frac {1}{t}, \quad w=\frac{{\mathfrak w}}{t^n} \quad \text{and, therefore},
\quad \xi = \frac{t^{n-1}}{{\mathfrak w}}, \quad \zeta=\frac{t^{n}}{{\mathfrak w}} \, .
\end{equation}
Substituting this into (\ref{curve}) and multiplying by $t^{2nr}$, we obtain the equation
$$ 
({\mathfrak w}^2-t)^r+ \text{higher order terms} =0 . 
$$ 
This shows that $\cal S$ has a strong singularity at the infinity.
To regularize $\cal S$ there, we construct the Puisaux expansions of ${\mathfrak w}$ in powers of $t$.
For this purpose, consider first the expansion of the polynomial Lax matrix \eqref{polynomial-L}
\begin{align}
{\bf L}(\la) & = \begin{pmatrix} {\mathcal V}(\la) & {\mathcal U}(\la) \\
    {\mathcal W}(\la) & -{\mathcal V}^T(\la) \end{pmatrix} \notag \\
& = \begin{pmatrix} \qquad \la^{n-1} {\mathcal V}_0 + \la^{n-2} {\mathcal V}_1  + \cdots &
\la^{n-1} {\bf I}_r  + \la^{n-2} {\mathcal U}_1 + \cdots \\
\la^{n} {\bf I}_r + \la^{n-1}{\mathcal W}_0 + \la^{n-2} {\mathcal W}_1 + \cdots &
- \la^{n-1} {\mathcal V}_0^T  - \la^{n-2} {\mathcal V}_1^T + \cdots \end{pmatrix}  \label{symp_form}
\end{align}
with the leading $r\times r$ matrix coefficients\footnote{We choose the indices of the coefficients in the same way as in \eqref{odd}}
$$
{\mathcal V}_0 = X^T P , \quad {\mathcal V}_1 = X^T A P , \quad
{\mathcal U}_1 = -\tr A \, {\bf I}_r + X^T A X , \quad  {\mathcal W}_0 = -\tr A\, {\bf I}_r - P^T P\, ,
$$
which are $r\times r$ matrix generalizations of the coefficients \eqref{u_w}.
Recall that the entries of ${\mathcal V}_0 = X^T P$ are first integrals of the system: ${\cal V}_0 =\frac 12 \Psi$.
Substituting \eqref{tW} into the eigenvector equation
$$
{\bf L}(\la)\psi =w\psi, \quad \psi\in {\mathbb P}^{2r-1},
$$
we get the expansion
\begin{equation}
\label{expansion}
\begin{pmatrix}  \qquad {\mathcal V}_0 t+ {\mathcal V}_1 t^2 + \cdots &
{\bf I}_r t  + {\mathcal U}_1 t^2 + \cdots \\
{\bf I}_r + {\mathcal W}_0  t + {\mathcal W}_1 t^2 + \cdots &
- {\mathcal V}_0^T t  -  {\mathcal V}_1^T t^2 + \cdots \end{pmatrix} \psi(t) = {\mathfrak w}(t) \psi(t).
\end{equation}
As $t \to 0$, the eigenvectors of ${\bf L}(\la)$ tend to those of $Y$ in \eqref{Nr}, which has $r$ different eigenvectors with zero eigenvalue.
As a result, the regularized curve ${\cal S}'$ must have $r$ different points over $\la =\infty$,
which are all ordinary branch points of the covering ${\cal S}'\to {\mathbb P}(\la)$,
and will be denoted as $\infty_1, \dots, \infty_r$. For a
local coordinate near each of these points we take $\tau=\sqrt{t}$, so that $\lambda=1/\tau^2$.

The structure of \eqref{expansion} allows to evaluate the
expansions of ${\mathfrak w}$, $\psi$ in the neighborhood of
each $\infty_s$, which will be used in the next sections.

\begin{prop} \label{puis}
\begin{description}
\item{1)}
Let rank of ${\mathcal V}_0 = X^T P$ be maximal
 and $r$ be odd. Then let
$$
\{ \nu_1,\dots \nu_{[r/2]}, -\nu_1,\dots -\nu_{[r/2]}, 0 \}
$$
be the set of the eigenvalues of the matrix coefficient ${\cal
V}_0\in \mathrm{so}(r)$ and
$$
\{ {\bf v}_1,\dots {\bf v}_{[r/2]},\, \bar{\bf v}_1,\dots,\bar{\bf v}_{[r/2]}, {\bf v}_0 \}
$$
be the corresponding eigenvectors normalized by
$$
\langle \beta, {\bf v}_s\rangle=1, \quad \langle \beta, \bar{\bf v}_s\rangle=1, \quad \langle \beta, {\bf v}_0\rangle=1,
$$
where $\beta=(\beta_1,\dots, \beta_r)\in {\mathbb C}^r$ is any non-zero vector,
not orthogonal to any eigenvector of ${\cal V}_0$.
Then in the neighborhood of $\infty_1,\dots, \infty_r$, the expansions ${\mathfrak w}(\tau)$ are, respectively,
\begin{align}
 {\mathfrak w}_s(\tau) & = \tau + \nu_s \tau^2 + b_s \tau^{3} +  \cdots, \notag \\
 {\mathfrak w}_{[r/2]+s} (\tau) & = \tau - \nu_s \tau^2 + b_s \tau^{3} -  \cdots, \qquad s=1,\dots, [r/2], \label{ast} \\
{\mathfrak w}_r(\tau) & = \tau \quad \quad \; + b_0 \tau^{3} + B_0 \tau^{5} + \cdots , \notag
\end{align}
and the corresponding expansions of the eigenvector
$\psi(\tau)=(\psi^1,\dots, \psi^{2r})^T$ normalized by
\begin{equation}\label{norm}
\beta_1 \psi^{r+1}+\cdots + \beta_r \psi^{2r}=1
\end{equation}
are
\begin{align}
\psi_s (\tau) & = \begin{pmatrix} {\bf v}_s \, \tau + {\bf h}_s \tau^2 + O(\tau^{3}) \\
{\bf v}_s + {\bf h}_s \tau + O(\tau^2) \end{pmatrix}, \notag \\
\psi_{[r/2]+s}(\tau) & = \begin{pmatrix} \bar{\bf v}_s \, \tau+ \bar {\bf h}_s \tau^2 + O(\tau^{3}) \\
  \bar{\bf v}_s + \bar{\bf h}_s \tau + O(\tau^2) \end{pmatrix},
\label{exp_psi} \\
\psi_r ( \tau) & = \begin{pmatrix} {\bf v}_0 \, \tau+ {\bf h}_0 \tau^2 + O(\tau^{3} ) \\
{\bf v}_0 + {\bf h}_0 \tau + O(\tau^2 ) \end{pmatrix}. \notag
\end{align}
Here $b_s, b_0, B_0$ are certain constants, and ${\bf h}_s, \bar{\bf h}_s, {\bf h}_0$
are constant vectors.
The last expansions ${\mathfrak w}_r (\tau), \psi_r (\tau)$ contain only odd powers of $\tau$.

\item{2)} If  rank of ${\mathcal V}_0$ is maximal and $r$ is even, then
item 1) holds without the last expansions ${\mathfrak w}_r(\tau)$, $\psi_r ( \tau)$.

\item{3)} If ${\mathcal V}_0 = 0$, then all the expansions contain only odd powers of $\tau$.\footnote{The spectral curve $\mathcal S$ is determined by the values of the commuting integrals
$\{{\cal I}_{2l}(\lambda)\}$ that also fix the values of the invariants $\tr(\Psi^{2k})$ and the rank of the momentum mapping $\Psi$.
Generically, $\mathrm{rank}(\Psi)=2[r/2]$. Here, by assuming $\Psi=\mathcal V_0=0$ we mean that we take appropriate
generic values of the integrals $\{{\cal I}_{2l}(\lambda)\}$ (in the real domain ensuring the complete integrability of the Neumann system on the Grassmannian variety $G_{n,r}$).}
\end{description}
\end{prop}

\noindent{\it Proof.} It is sufficient to substitute the above
expansions into \eqref{expansion} and compare few leading
terms in both sides. \hfill$\square$

\medskip

The involution $\sigma(\lambda,w)=(\lambda,-w)$ can be naturally continued to the involution on the regularized curve $\mathcal S'$
by using the mapping
\begin{equation}\label{sigma3}
\tau \longmapsto -\tau.
\end{equation}

Namely, if $\mathrm{rank}(\mathcal V_0)=2[r/2]$, \eqref{sigma3}  transforms
the series ${\mathfrak w}_s(\tau)$, $s=1,\dots, [r/2]$ in \eqref{ast} to
$-{\mathfrak w}_{[r/2]+s}(\tau)$ and vice versa  (if $r$ is odd, $\mathfrak w_r(-\tau)=-\mathfrak w_r(\tau)$). Since
$\lambda(\tau)=1/\tau^2$ and $w(\tau)=\mathfrak w(\tau)/\tau^{2n}$, by taking the appropriate branches
of the coverings ${\cal S}'\to {\mathbb P}(\la)$ and  ${\cal S}\to {\mathbb P}(\la)$ (over a neighborhood of $\lambda=\infty$), we get that the mapping \eqref{sigma3} without $\infty_1,\dots,\infty_r\in\mathcal S'$ coincides with \eqref{sigma} without the singular point $(0:1:0)\in\mathcal S$.
It follows that the infinite points of ${\cal S}'$ admit the division into the subsets
\begin{equation} \label{inf_points}
\begin{aligned}
\{\infty_1^-,\dots \infty_{r/2}^-\}, & \quad \{\infty_1^+,\dots \infty_{r/2}^+\}  \qquad\qquad\qquad  \text{if $r$ is even},\\
\{\infty_1^-,\dots \infty_{[r/2]}^-\}, &\quad \{\infty_1^+,\dots
\infty_{[r/2]}^+\}, \quad \{\infty_0\} \qquad \text{if $r$ is odd},
\end{aligned}
\end{equation}
such that the extension of $\sigma$ is defined by
\begin{equation}\label{sigma2}
\sigma ( \infty_j^\mp )=\infty_j^\pm, \qquad j=1,\dots, [r/2], \qquad \sigma(\infty_0)=\infty_0
\end{equation}
and \eqref{sigma3} maps the neighborhoods of $\infty^-_j,\infty^+_j,\infty_0$ to the neighborhoods of $\infty^+_j$, $\infty^-_j$, $\infty_0$, respectively.

If ${\mathcal V}_0 = 0$,  from item 3) of Proposition \ref{puis}, all the infinite points are invariant with respect to the involution:
\begin{equation}\label{sigma4}
\sigma(\infty_j)=\infty_j, \qquad j=1,\dots,r.
\end{equation}

Using the expansions ${\mathfrak w}_s(t)$ and the second pair of relations (\ref{tW}),
one calculates the set of the Puisaux expansions of the coordinate $\zeta$ in powers of $\xi$ in the neighborhood of $(0:1:0)$.
Then, applying the theory of singularities of algebraic curves (see e.g., \cite{Casas, Kir}),
we obtain the order $\delta_\infty$ of singularity of $\cal S$ at
$(0:1:0)$\footnote{These long calculations are due to Maria Alberich (UPC)}:
\begin{equation} \label{singul_order}
\begin{aligned}
\delta_\infty & = 2nr(nr-2r-1)+2r(r+1) \qquad \qquad \text{if rank$(X^T P)=2[r/2]$},  \\
\delta_\infty & = 2nr(nr-2r-1)+r^2 + \frac{3r(r+1)}{2} \qquad \text{if $X^T P=0$.}
\end{aligned}
\end{equation}

Now, summing the singularity orders $\delta_\infty$ and
$\delta_1,\dots, \delta_n$ and taking into account that the degree
of the curve $\cal S$ equals
$$
\mathrm{deg}(\mathcal S)= n(2r-1)+(n-r)=2nr-r,
$$
we use the Pl\"ucker formula to get the following result.

\begin{thm} \label{big_g} For generic values of commuting integrals \eqref{perelomov},
the geometric genus of $\cal S$ equals
\begin{equation*}
g  =  \frac{(\mathrm{deg}(\mathcal S)-1)(\mathrm{deg}(\mathcal S)-2)}{2}- \sum_{i=1}^n \delta_i -\delta_\infty =
2nr-n - \frac 32 r^2- \frac 12 r+1.
\end{equation*}
If $\Psi=0$ we have
$g =2r(n-r) -n+1$.
\end{thm}


\subsection{The ''small'' curve ${\cal C}= {\cal S}/\sigma$ and its genus.}
In view of the involution \eqref{sigma}, the curve $\cal S$ is a
2-fold ramified covering of the curve $ {\mathcal C} \subset
{\mathbb C}^2(u,\la)$, $u=w^2$, given by the equation
$$
{\cal C}\colon \quad u^{r} + u^{r-1} a(\lambda){\cal I}_{2}(\lambda)+ \cdots
+ u\cdot a^{2r-3}(\lambda) {\cal I}_{2r-2}(\lambda)+ {a}^{2r-1} (\lambda){\cal I}_{2r}(\lambda)=0
$$

Making the birational transformation $(\la, u)\to (\la, \mu)$ with
$u= a(\la)\, \mu$, we see that the equation of $\cal C$ gives the
lines $\{\la=a_i\}$ and the curve
\begin{equation} \label{eq_C}
\widetilde{\mathcal C} \colon \quad \mu^{r} + \mu^{r-1} {\cal I}_{2}(\lambda)+ \cdots + \mu\, a^{r-2}(\lambda) {\cal I}_{2r-2}(\lambda)
+ {a}^{r-1} (\lambda){\cal I}_{2r}(\lambda)=0 .
\end{equation}
The latter is singular over $\lambda=a_1, \dots, a_n$ and at its infinite part.
Let ${\mathcal C}'$ be a compete regularization of $\widetilde{\mathcal C}$.
Then the regularized curve ${\mathcal S}'$ is a 2-fold covering of $\mathcal C'$.

The covering
\begin{equation}\label{covering}
\pi\colon\, {\mathcal S}'\longrightarrow {\mathcal C}'=\mathcal S'/\sigma
\end{equation}
 is branched when $\mu=0$,
that is, when $\la=a_i$ or $\la$ is a simple root of the last polynomial
${\cal I}_{2r}(\lambda)$ of degree $n-r$ in \eqref{eq_C}.
As mentioned above, the projection ${\cal S}'\to {\mathbb P}\{\la \}$ has only one ordinary branch point $P_{a_i}$ over
$\la=a_i$, hence $P_{a_i}$ is also the only branch point of \eqref{covering} over $\la=a_i$.
(This means that the projection ${\mathcal C}'\to {\mathbb P}\{\la \}$ is not ramified over $\la = a_i$.)
The roots of ${\cal I}_{2r}(\lambda)$ give other $n-r$ ordinary branch points of $\pi$.
Therefore, $\pi$ always has $2n-r$ {\it finite} ordinary branch points.

Next, as follows from \eqref{sigma2}
if rank$(\Psi)=2[r/2]$ and $r$ is odd, the covering \eqref{covering} is also ramified at $\infty_0$,
and it is not ramified over infinity when $r$ is even.

If $\Psi=0$, \eqref{sigma4}
implies that $\infty_1,\dots,\infty_r$ are ordinary
branch points of $\pi$.

As a result, the covering \eqref{covering} has in total
$2n-2[r/2]$ ordinary branch points if rank$(\Psi)=2[r/2]$, and $2n$ ordinary branch points if $\Psi =0$.
Then, according to the Riemann--Hurwitz formula,
\begin{equation} \label{R-H}
\begin{aligned}
g= \text{gen}({\mathcal S}' ) & = 2( \mbox{gen}({\mathcal C}') -1)+
\frac {2n-2[r/2]}2+1,  \quad \text{if rank$(\Psi)=2[r/2]$,} \\
 g & = 2( \mbox{gen}({\mathcal C}') -1)+ \frac {2n}{2} +1,  \quad \qquad \qquad  \text{if $\Psi = 0$}.
\end{aligned}
\end{equation}
In view of Theorem \ref{big_g}, this gives

\begin{prop} \label{gen_g_0} If {\rm rank}$(\Psi)=2[r/2]$,
\begin{equation*}
g_0 =\mathrm{gen}({\cal C}') = \left\{ \begin{aligned} & n(r-1)- 3 (r^2-1)/4 \quad \mbox{$r$ is odd}, \\
  & n(r-1)-3 r^2/4 + 1 \qquad \mbox{$r$ is even}   \end{aligned} \right.
\end{equation*}
and, if $\Psi = 0$, $g_0 = r(n-r) + n-1$.
\end{prop}

In the next section the genera of ${\cal S}', {\cal C}'$ will be used to calculate the dimension of
the Abelian subvarieties of the
Jacobian of ${\cal S}'$, which are related to complex invariant tori of the system.

\paragraph{The case $r=2$.} In this simplest case it is possible to calculate the genera $g_0, g$ without long calculations based on the above Puisaux expansions. Namely, the curve $\tilde{\mathcal C}$ in \eqref{eq_C} takes the simple form
$$
\tilde{\mathcal C}\colon\quad \mu^{2} + \mu \, {\cal I}_{2}(\lambda) + a (\lambda) {\cal I}_{4}(\lambda)=0 .
$$
which, under the birational change $\mu= -{\cal I}_{2}(\lambda)/2 + y/2$,
gives the Weierstrass hyperelliptic form
\begin{equation} \label{ground}
y^2= {\cal I}_{2}^2(\lambda)- 4 a (\lambda) \, {\cal I}_{4}(\lambda ).
\end{equation}
To find its genus, note that
here ${\cal I}_{2}(\lambda), {\cal I}_{4}(\lambda )$ are polynomials of degrees $n-1, n-2$ respectively, hence one
might expect
the right hand side of (\ref{ground}) to be a polynomial of degree $2n-2$.
However, in the general case (when $\Psi \ne 0$), the degree is
$$
\mathrm{deg}(\tilde{\mathcal C})=2n-3=2(n-2)+1
$$
(in view of the expressions
for ${\cal I}_4(\la), {\cal I}_2(\la)$, the coefficient at the leading power $\la^{2n-2}$ vanishes).
Then the regularization ${\mathcal C}'$ of $\tilde{\mathcal C}$ has genus $g_0=n-2$ and one infinite point.

Next, as was counted above, for $r=2$ and $\Psi\ne 0$, the
covering \eqref{covering} has in total
$2n-2$ ordinary ramification points. Hence, by the
Riemann--Hurwitz formula \eqref{R-H},  the genus of the ''big''
curve ${\cal S}'$ is
$$
g= 2(n-2-1)+(2n-2)/2+1 = 3n-6 \, ,
$$
and ${\cal S}'$ has 2 infinite points over $\lambda=\infty$, which pass to each other under $\sigma$.

In the case $\Psi=0$ the right hand side of (\ref{ground}) becomes a polynomial of degree $2n-4$, the curve
${\mathcal C}'$ has now 2 infinite points,
its genus drops to $g_0-3$, whereas the covering $\pi$ has $2n$ ordinary ramifications. All this
yields $g=3n-7$.


\section{Complex invariant manifolds}

From the curve ${\cal S}'$ we pass to its Jacobian variety,
$\Jac({\mathcal S}')$, defined as the additive group of degree
zero divisors on ${\cal S}'$ modulo divisors of meromorphic
functions. Equivalently, in some cases we will consider effective
divisors $D$ as elements of $\Jac({\mathcal S}')$ by choosing a
basepoint $P_0\in {\cal S}'$ and associating to $D$ the degree
zero divisor $D- N P_0$, where $N$ is the degree of $D$.

\subsection{The Prym variety}
The involution $\sigma: {\mathcal S}' \to {\mathcal S}'$,
$\sigma(\la,w)\to (\la, -w)$ extends to the Jacobian of ${\mathcal
S}'$. Hence $\Jac({\mathcal S}')$ contains two Abelian
subvarieties: the Jacobian of the ''small'' underlying curve
${\mathcal C}'$ of dimension $g_0=\mathrm{gen}({\mathcal C}')$ and
$\Prym ({\mathcal S}',\sigma)$ of dimension
$$
\dim(\Prym ({\mathcal S}',\sigma))=
\text{gen}({\mathcal S}' )-\text{gen}( {\mathcal C}' ).
$$
The
vectors of $\Jac ({\mathcal C}') \subset \Jac ({\mathcal S}') $
are invariant with respect to $\sigma$, whereas the vectors of
$\Prym ({\mathcal S}',\sigma) \subset\Jac ({\mathcal S}')$ are
anti-invariant.

Algebraically, $\Prym ({\mathcal S}',\sigma)$ is defined as the
set of degree zero divisors $\cal D$ on $\mathcal S'$ such that
${\cal D} + \sigma {\cal D}$ is the divisor of a meromorphic
function on $S'$. For a given degree zero divisor $\cal H$, a
translate of $\Prym ( {\mathcal S}',\sigma)$ in $\Jac({\mathcal
S}')$ is the set of degree zero divisors $\cal D$ satisfying
$$
{\cal D} + \sigma {\cal D} \equiv {\cal H} ,
$$
where $\equiv $ denotes the equivalence modulo divisors of meromorphic functions on $S'$.

The variety $\Jac ({\mathcal S}')$ is isogeneous (but not isomorphic) to
$\Jac ({\mathcal C}')\times \Prym ({\mathcal S}', \sigma)$. A detailed algebraic description of
Prym varieties in the case of a general double covering ${\cal S}'\to {\cal C}'$ ramified at $N>2$ points is given in
\cite{Fay} (Chapter V). Using this description and our previous calculations, one can show that
$\Prym ({\mathcal S}', \sigma)$ is an Abelian variety with polarization
$$
(\underbrace{1,\dots,1}_{g_0}, \underbrace{2\dots,2}_{n-[r/2]-1}), \qquad (\underbrace{1,\dots,1}_{g_0}, \underbrace{2\dots,2}_{n-1}),
$$
for $\Psi$ of maximal rank and  $\Psi=0$, respectively. We will not use the polarization data in the sequel.

From Theorem \ref{big_g} and Proposition \ref{gen_g_0} we immediately obtain
\begin{prop} \label{dims}
\begin{description}
\item{1)} For generic constants of motion, the
dimension of the Abelian subvariety $\Prym ({\mathcal S}', \sigma)$ equals
$$
l= \frac 12 \big(2 r(n-r)+\frac{r(r-1)}2-\left[\frac{r}{2}\right]\big),
$$
which coincides with one half of the dimension of a generic symplectic leaf ${\cal U}_c$ in $T^*V_{n,r}/SO(r)$.

\item{2)} In the special case $\Psi=2 X^T P=0$ the dimension of
$\Prym ({\mathcal S}', \sigma)$ is $r(n-r)$,
which coincides with the dimension of $G_{n,r}$.
\end{description}
\end{prop}

Note that in the simplest case $r=2$ (and only in this case), the
dimension of $\Prym ({\mathcal S}', \sigma)$ equals $2(n-2)$,
i.e., the dimension of the Grassmanian $G_{n,2}$, regardless to
whether $\Psi=0$ or not.

\subsection{The generalized Jacobian and the affine Prym subvariety.}

Assuming that the rank of the momentum $\Psi$ is maximal, we also
introduce the generalized Jacobian of the singularized curve
${\mathcal S}''$ obtained from ${\mathcal S}'$ by gluing pairwise the
involutive infinite points $\infty_s^-, \infty_s^+$,
$s=1,\dots,[r/2]$ described in \eqref{inf_points}, \eqref{sigma2}. Namely
$$
\mathcal S''=\mathcal S_0\cup \{ \widehat\infty_1,\dots,
\widehat\infty_{[r/2]} \}, \quad \mathcal S_0=\mathcal S'\setminus
\{ \infty_1^-,\infty_1^+, \dots, \infty_{[r/2]}^-,\infty_{[r/2]}^+
\}
$$
so that $\widehat\infty_1,\dots, \widehat\infty_{[r/2]}$ are
ordinary double points of $S''$. Then a meromorphic function $f$
on $\mathcal S'$ will be also meromorphic on $\mathcal S''$ if
$f(\infty_s^-)=f(\infty_s^+)$.

The generalized Jacobian of $\mathcal S''$, denoted as
$\widetilde{\Jac}({\mathcal S}', \infty)$ is then defined as the
set of degree zero divisors $\cal P$ on the affine part $\mathcal
S_0$ considered modulo $\infty$-equivalence: ${\cal P}
\equiv_\infty {\cal Q}$ if there exists a meromorphic function
$f$ on the smooth curve $S'$ such that
$$
1)\;  (f) = P-Q\, ,\quad 2) \; f(\infty_s^-)=f(\infty_s^+) \ne 0, \infty , \qquad s=1,\dots,[r/2]
$$
(note that the values $f(\infty_i^\pm), f(\infty_j^\pm)$ for $i\ne
j$ can be different).\footnote{The above definition should be
distinguished from the other kind of generalized Jacobians, when
several points of a smooth curve are glued to the same point, as
described in \cite{Beau, Serre, Gavr}. Both kinds include the
simplest case when the singularized curve has only one double
point.}

Thus, there is an exact sequence of algebraic groups
\begin{equation} \label{ex_sec}
\begin{CD}
 0 @ > \exp >> ({\mathbb C}^*) ^{[r/2]} @> \upsilon >>  \widetilde{\Jac}(\mathcal S', \infty) @ > \phi >> \Jac ({\mathcal S}')
 @ > >> 0
\end{CD}
\end{equation}
where, for $\rho=(\rho_1,\dots,\rho_{[r/2]})\in {\mathbb C^*}^{[r/2]}$, the image $\upsilon(\rho)$ is the divisor
of any meromorphic function $f(p)$ on $S'$ satisfying
\begin{equation} \label{quo_rho}
f(\infty_s^+)/f(\infty_s^-)=\rho_s,   \qquad s=1,\dots,[r/2],
\end{equation}
and for any degree zero divisor $\cal P$,  $\phi({\cal P})$ gives a point in $\Jac ({\mathcal S}')$.

The analytical description of $\widetilde{\Jac}({\mathcal S}', \infty)$ is following:
let $\omega_1,\dots,\omega_{g}$ be $g$ independent holomorphic differentials on $S'$
and $\varOmega_{{j}}$, $j=1,\dots,[r/2]$ be meromorphic differential of the 3rd
kind having a pair of simple poles at $\infty_j^+, \infty_j^-$ respectively.
One can always normalize the meromorphic differentials in such a way that they will be anti-invariant under the involution:
$\sigma^* \varOmega_{{j}}= - \varOmega_{{j}}$.

Next, let $\varLambda\subset {\mathbb C}^{g} (z_1,\dots, z_g)$ be
the lattice generated by vectors of periods of
$(\omega_1,\dots,\omega_{g})^T$ with respect to a basis in
$H_1({\cal S}',{\mathbb Z})$. Respectively, let
$$
\widetilde\varLambda\subset {\mathbb C}^{g+[r/2]} (z_1,\dots,
z_g,Z_1,\dots,Z_{[r/2]})
$$
be the lattice generated by vectors of periods of $
(\omega_1,\dots,\omega_{g}, \varOmega_{1},\dots,
\varOmega_{[r/2]})^T $ with respect to a basis in $H_1({\cal
S}_0,{\mathbb Z})$ formed by $2g$ canonical cycles on $S'$ and the
homology zero cycles embracing $\infty_1^+,\dots,
\infty_{[r/2]}^+$. The lattice $\widetilde\varLambda$ has rank
$2g+[r/2]$, and the generalized Jacobian
$\widetilde{\Jac}({\mathcal S}', \infty)$ is the factor ${\mathbb
C}^{g+[r/2]}/\widetilde\varLambda$, which is a non-compact
algebraic group.
The map $\phi$ in \eqref{ex_sec} acts as projection:
\begin{equation}\label{projection}
\phi (z_1,\dots, z_g,Z_1,\dots,Z_{[r/2]}) =(z_1,\dots, z_g).
\end{equation}

The relation between the above analytical and algebraic
descriptions of $\widetilde{\Jac}({\mathcal S}', \infty)$ is given
by the generalized Abel map with a basepoint $P_0 \in S_0$
\begin{equation}\label{GAM}
\tilde{\cal A} (P) =\int_{P_0}^{P} \left(
\omega_1,\dots,\omega_{g}, \varOmega_{1},\dots, \varOmega_{[r/2]}
\right)^T \in {\mathbb C}^{g+[r/2]}\, , \quad P \in S_0.
\end{equation}
A general point in  $\widetilde{\Jac}( {\mathcal S}', \infty)$ is the Abel image of an effective divisor of
degree $\ge g+[r/2]$ on ${\mathcal S}_0$.
The inversion of the generalized Abel map in terms of generalized theta-functions
is described in \cite{Cl_G, Gen_Jac}.

A generalization of the Abel theorem says that two effective
divisors
$$
\mathcal P=P_1+\dots+P_N, \qquad \mathcal Q=Q_1+\dots+Q_N
$$
on
${\mathcal S}_0$ are $\infty$-equivalent ($\mathcal P\equiv_\infty \mathcal Q$) if and only if
$$
\tilde{\cal A} (P_1)+ \cdots + \tilde{\cal A} (P_N)= \tilde{\cal
A} (Q_1)+ \cdots +\tilde{\cal A} (Q_N)   \quad \text{modulo $\widetilde\varLambda$}.
$$
Note that 
two divisors ${\cal D}_1, {\cal D}_2$ on ${\mathcal S}_0$ can correspond to the same point on
$\Jac ({\cal S})$, but to different points on $ \widetilde {\Jac}( {\mathcal S}', \infty)$.

\paragraph{The affine Prym variety.}
The involution \eqref{sigma} extends to $\widetilde \Jac(
{\mathcal S}',\infty)$, which then contains two algebraic
subgroups: one is invariant with respect to $\sigma$ and the other
is anti-invariant. The first is just the usual Jacobian $\Jac
({\mathcal C}')$ of dimension $g_0$. The second is
 the {\it affine Prym subvariety}\footnote{We do not use term "generalized Prym variety", quite natural in this situation,
since it is related to another construction considered in
\cite{be77}.} $\widetilde \Prym(\mathcal S',\sigma)\subset
\widetilde \Jac(\mathcal S',\infty)$, which can be defined as the
set of degree zero divisors $\cal P$ on ${\mathcal S}_0$ such that
\begin{align*}
  {\cal P} + \sigma {\cal P} & = \{\text{a divisor of a meromorphic function $f$} \\
& \qquad \text{ with $f(\infty_s^+)=f(\infty_s^-) \ne 0, \infty, \quad s=1,\dots,[r/2]$} \;  \} \\
& = \text{the origin in $\widetilde \Jac(\mathcal S',\infty)$ }.
\end{align*}

In view of the definition of the map $\upsilon \, :\, ({\mathbb
C}^*) ^{[r/2]} \mapsto \widetilde{\Jac}(\mathcal S', \infty)$, the
action of $\sigma$ on the set $\upsilon \left( ({\mathbb
C}^*)^{[r/2]} \right)$ is given by
$$
\sigma (\upsilon( \rho_1,\dots,\rho_{[r/2]} )) =\upsilon(1/\rho_1,\dots,1/\rho_{[r/2]} )  \, .
$$
Then we obtain

\begin{lem} \label{lem_ups}
The set $\upsilon \left( ({\mathbb C}^*)^{[r/2]} \right)$ belongs
to $\widetilde \Prym(\mathcal S',\sigma)$. The involution $\sigma$
on it has $2^{[r/2]}$ fixed points $\upsilon(\pm 1,\dots,\pm 1)$,
which are also half-periods in $\widetilde \Jac(\mathcal
S',\infty)$ and in $\widetilde \Prym(\mathcal S',\sigma)$.
\end{lem}

\noindent{\it Proof.} Let ${\cal P}$ be the divisor of any
meromorphic function $f(p)$ on $\mathcal S'$ satisfying
\eqref{quo_rho}, and let $f_\sigma (p)=f(\sigma(p))$ so that
$f_\sigma (\infty^+_s)/f_\sigma(\infty^-_s)=1/\rho_s$. Then
$$
(f\cdot f_\sigma)= {\cal P} + \sigma {\cal P}, \qquad f(\infty^+_s)
f_\sigma (\infty^+_s)= f(\infty^-_s) f_\sigma (\infty^-_s),
$$
hence, by the definition, $\cal P$ belongs to $\widetilde
\Prym(\mathcal S',\sigma)$.

Next, if ${\cal P}$ is the divisor of a meromorphic function $f(p)$ satisfying $f(\infty^+_s)=\pm f(\infty^-_s)$, then
 $(f^2)= 2{\cal P}$ and $f^2(\infty^+_s)= f^2(\infty^-_s)$. That is,
$2 {\cal P}$ corresponds to the origin in $\widetilde{\Jac}(S',
\infty)$, and ${\cal P}$ is a half-period of the generalized
Jacobian and of the affine Prym subvariety. \hfill$\square$

\medskip

By the above lemma, the exact sequence \eqref{ex_sec} implies the following sequence of groups
$$
\begin{CD}
0 @> \exp >> ({\mathbb C}^*)^{[r/2]}  @> \upsilon >> \widetilde\Prym(S',\sigma) @ > \phi >>
\Prym(S,\sigma) @ > >> 0  .
\end{CD}
$$
Since $\phi(\upsilon \left( ({\mathbb C}^*)^{[r/2]} \right) )=0$, the sequence is also exact.

It follows that, like $\widetilde \Jac( {\mathcal S}',\infty)$, the variety $\widetilde\Prym ({\mathcal S}', \sigma)$
is non-compact and
has dimension $g+[r/2]-g_0=l+[r/2]$, which, in view of Proposition \ref{dims}, equals $\delta$,
the dimension of generic invariant tori in $T^*V_{n,r}$.

The generalized Jacobian $\widetilde \Jac( {\mathcal S}',\infty)$ is isogeneous (but not isomorphic) to the product
$\Jac({\mathcal C})\times \widetilde\Prym ({\mathcal S}', \sigma)$.

\subsection{The eigenvector map}\label{EVM}

To proceed further, we need to recall some basic notions relating
Lax matrices and Abelian varieties, and we follow the description
of \cite{Dub_0, Krich, AHH2, Beau, RS}.

Let ${\mathbb L}(\la)$ be a $d\times d$ polynomial Lax matrix and $S$ be its spectral curve
(with its infinite points), which is assumed to be regular or appropriately regularized, of geometric genus $g$.
${\mathbb L} (\la)$ defines the eigenvector bundle
${\varepsilon }\, : {S} \longrightarrow {\mathbb P}^{d-1}$ given by the eigenvectors
$\psi (P)=( \psi^1(P), \dots,\psi^{d}(P))^T $, $P\in {S}$ of ${\mathbb L} (\la)$.

Impose a normalization $\langle \alpha, \psi (P) \rangle=1$, where
${\alpha}=(\alpha_1,\dots,\alpha_{2r})^T\in {\mathbb P}^{2r-1}$ is an arbitrary
constant vector, and
consider the {\it minimal} effective divisor ${\cal D}_\alpha$ on ${S}$ such that
$$
\big( \psi^{l}(P) \big) = \text{zeros of $\psi^{l}(P)$
  - poles of $\psi^{l}(P)$} \ge - {\cal D}_\alpha , \quad l=1\dots,d,
$$
that is, each of the normalized components $\psi^{l}(P)$ can have poles {\it at most} at the points of ${\cal D}_\alpha$.
These points can be found as common zeros of certain polynomials of $\la,w$ (see, e.g., \cite{AHH2}). Also
\begin{equation} \label{deg}
\text{deg} ({\cal D}_\alpha) = g+d-1 =N .
\end{equation}
(One can always chose such a normalization $\alpha$
that $d-1$ points of ${\cal D}_\alpha$ will be fixed in the infinite part of ${S}$.)

It is known that two divisors ${\cal D}_\alpha$, ${\cal
D}_{\alpha'}$ corresponding to different normalizations
$\alpha,\alpha'$ are linearly equivalent. Therefore, for a
basepoint $P_0\in S$, the degree zero divisors ${\cal D}_\alpha -
N P_0, {\cal D}_{\alpha'}-N P_0$ give the same point in $\Jac
(S)$, more precisely, in the open subset $\Jac
(S)\setminus\Theta$, with $\Theta\in \Jac (S)$ being a translate
of the theta-divisor, the Abel image of all the special divisors
on $S$.

Let now ${\cal I}_S$ be {\it the isospectral manifold}: the set of all the above matrices ${\mathbb L}(\la)$ having the same
spectral curve $S$. Thus we get {\it the eigenvector map} ${\cal E}: \; {\cal I}_S \to \Jac (S)$.

If the equivalence
class $\{{\cal D}\}$ is the image of a matrix ${\mathbb L}(\la)\in {\cal I}_S$, the latter can be reconstructed up to a conjugation
by an element of the group $\mathbb {PGL}(d,{\mathbb C})$ (not depending of $\la$).  The main steps of this are as follows: Let $L({\cal D})$ be the
vector space of meromorphic functions $f(P)$ on $S$ with $(f)\ge {\cal D}$ (it includes $f\equiv 1$). According to the
Riemann--Roch theorem, for a non-special degree $N$ divisor ${\cal D}$, dim $(L({\cal D}))=N-g+1$, hence, in the considered
case, dim $(L({\cal D}))$ is precisely $d$.

There exists a basis $\{f_1(P),\dots,f_d (P)\}$ in $L({\cal D})$
such that ${\bf f}(P)=(f_1,\dots,f_d)^T$, $P=(\la,\mu)\in S$ is an
eigenvector\footnote{It is always supposed that ${\bf f}$ is
normalized, i.e., its components do no have a common zero} of
${\mathbb L}(\la)$. To reconstruct ${\mathbb L}(\la)$, choose
$\la\in {\mathbb C}$ with distinct eigenvalues $\mu_1,\dots,\mu_d$
and consider the $d\times d$ matrix
$$
\mathbf F (\la) =( {\bf f} (P_1) \, \cdots \, {\bf f} (P_d)),
\qquad P_j=(\la,\mu_j).
$$
Then the matrix
\begin{equation} \label{reconstr}
   {\cal X} (\la) = \mathbf F(\la)\, \diag(\mu_1,\dots,\mu_d ) \, \mathbf F^{-1} (\la)
\end{equation}
is independent of the order of $\mu_1,\dots,\mu_d$, is, in fact, polynomial in $\lambda$,
and has the prescribed spectral curve $S$. It reconstructs the matrix ${\mathbb L}(\la)$ up to a
conjugation by a constant matrix (see e.g., \cite{Dub_0, AHH1}).

Clearly, two matrices ${\mathbb L}(\la), \tilde{\mathbb L}(\la)
\in {\cal I}_S$ that are conjugated by a constant element of
$\mathbb{PGL}(d,{\mathbb C})$ have equivalent divisors ${\cal
D},\tilde {\cal D}$. The inverse is also true: ${\cal D}\equiv
\tilde {\cal D}$ means the existence of a meromorphic function
$g(P)$ on $S$ with $(g)={\cal D}-\tilde {\cal D}$. Let the
components of $ {\bf f}(P), \tilde {\bf f}(P) $ form bases of
$L({\cal D}), L(\tilde {\cal D})$ respectively. Then the
components of $g(P) \tilde {\bf f}(P)$ form a basis of $L({\cal
D})$, and there is a non-degenerate matrix $R\in GL(d,{\mathbb
C})$ independent of $P$ such that ${\bf f}(P)= g(P) R \, \tilde
{\bf f}(P)$.

It follows that the induced map
$$
{\mathcal M}\colon
\{ {\mathbb L}(\la) \in {\cal I}_S \; \text{up to conjugation by matrices of}\; {\mathbb{PGL}}(d,{\mathbb C}) \} \mapsto \Jac(S)\setminus \Theta
$$
is injective.

When the leading coefficient of ${\mathbb L}(\la)$ is a constant matrix $J$
{\it and there are no constraints on the other coefficients},
it is natural to replace the isospectral manifold ${\cal I}_S$ by $\mathcal I_S^J$, the set of the matrices ${\mathbb L}(\la)$ having the same spectral curve $S$ and leading coefficient $J$.
In the general important case, when $J$ is diagonalizable, its stabilizer is the product
$ \left({\mathbb C}^*\right)^{d-1}$. Then, as was shown in many publications (see \cite{Dub_0, Beau, RS, AHH1}),
if the curve is smooth, the induced eigenvector map
$  {\cal I}_S^J/\left({\mathbb C}^*\right)^{d-1} \mapsto \Jac(S) \setminus \Theta$
is an isomorphism. The inverse map is described explicitly by
means of theta-functions associated to the curve ${S}$.

If $J$ in not diagonalizable and/or ${\mathbb L}(\la)$ belongs to a subgroup of $\widetilde{\mathrm {gl}}(d,{\mathbb C})$, the above result should be refined (one must
consider the equivalence by conjugations by elements of a subgroup of ${\mathbb{PGL}}(d,{\mathbb C})$).
Some cases of that have been considered in \cite{GG, Viv, Dra}.

In addition, some general properties of the eigenvector map in our case, when ${\mathbb L}(\la)$
belongs to the loop subalgebra
$\widetilde{\mathrm{sp}}(2r,{\mathbb C})$, were already studied in \cite{AHH1} (Section 5, case {\it ii}).

\subsection{The complex manifolds ${\cal I}, {\cal I}_{red}, {\cal I}_h$.}

Now consider the Neumann system on the complexified cotangent bundle
$T^*_{\mathbb C} V_{n,r}$ defined by the constrains
\eqref{cond_XP} in the complex domain. Let us fix generic values
of all commuting integrals $f_{k,i}$
described in Section \ref{NSV}. This also fixes
the values of the invariants $\tr(\Psi^{2j})$ and
the spectral curve
${\mathcal S}'$ of the "small" $2r\times 2r$ Lax matrix
\eqref{Lax_r}, \eqref{polynomial-L}, which belongs to the loop
subalgebra $\widetilde{sp}(2r,{\mathbb C})$.

Since there are $l+[r/2]$ independent commuting integrals, the corresponding invariant manifold
$$
\mathcal I=\{(X,P)\, \vert\,  f_{k,i}=c_{k,i}, \, \tr(\Psi^{2j})=c_j\}  \subset T^*_{\mathbb C}V_{n,r}
$$
has the (complex) dimension\footnote{In the real domain, $\mathcal I$ is a coisotropic invariant manifold  of our system.}
$$
\delta +r(r-1)/2 -[r/2]= l+ \text{dim}\; SO(r)
$$
with $\delta$ given by \eqref{dim_fol}.

The manifold $\cal I$ and the curve ${\cal S}'$ are invariant with
respect to the right action of the complex group $SO(r,{\mathbb
C}) $ on $(X,P)$, as well as to the action of the group ${\mathbb
Z}_2^n$ of reflections \eqref{reflection}. Moreover, the product
$SO(r,{\mathbb C})\times {{\mathbb Z}_2^n}$ is the maximal group
leaving $\cal I$ invariant. The reduced manifold ${\cal I}_{\text
{red}}= {\cal I} /SO(r,{\mathbb C})/{{\mathbb Z}_2^n}$ is the
complex extension of the corresponding $l$-dimensional invariant
isotropic torus  of the reduced Neumann system on
$T^*V(n,r)/SO(r)/\mathbb Z_2^n$.
Next, $\cal I$ itself is foliated by  complexified
$\delta$--dimensional invariant isotropic tori
$$
{\cal I}_h = {\cal I} \cap \Psi^{-1} (h)= \{ (X,P)\in {\cal I} \mid \Psi = h \},
$$
the joint level varieties of the extra non-commuting integrals of the set
$\{\Psi_{ij}\}$\footnote{In the simplest non-trivial case $r=2$ the manifold ${\cal I}_h$ coincides with $\cal I$.}.
Here $h\in \mathrm{so}(r,\mathbb{C})$ belongs to the adjoint orbit $\{\tr(\Psi^{2j})=c_j\}$.
As follows from the above, the factor of each ${\cal I}_h$ by the
direct product $({\mathbb C}^*)^{[r/2]}$, which is the complex
stabilizer of the momentum $h\in \mathrm{so}(r, {\mathbb C})$ in
$SO(r,{\mathbb C})$, and by the group of reflections ${\mathbb
Z}_2^n$ coincides with ${\cal I}_{\text{red}}$. The
relations between the manifolds can be described by the following
diagram showing the factorizations and the inclusion
\begin{equation*}
\xymatrix@R35pt@C50pt{
\mathcal I_h \ar@{^{(}->}[r]^{\Psi_{ij}=h_{ij}} \ar[dr]_{/({\mathbb C}^*)^{[r/2]}/{{\mathbb Z}_2^n }} &
\mathcal I \ar[d]^{/ SO(r,{\mathbb C})/{{\mathbb Z}_2^n }} \\
& \mathcal I_{\mathrm{red}} }
\end{equation*}

This can also be seen as follows. Note that
$\mathcal I/\mathbb Z_2^n$ can be identified with the
isospectral manifold\footnote{Actually $\mathcal I/\mathbb Z_2^n$ is a subset of the isospectral manifold ${\mathcal I}_{S'}$ of the curve $S'$
in a sense of a general definition given in the subsection \ref{EVM} (the $\mathbb{PGL}(2r,{\mathbb C})$--action does not leave it invariant).
However, as in the case when the leading term $J$ of the Lax matrix is fixed
where one use $\mathcal I^J_\mathcal S$ instead $\mathcal I_S$ (see the subsection \ref{EVM}), in our case
it is natural to take $\mathcal I/\mathbb Z_2^n$ instead ${\mathcal I}_{S'}$.}
 by assigning to each point $(X,P)$ the Lax matrix $\mathbf L(\lambda)$
in \eqref{symp_form}.
The conjugations that preserve $\mathcal I/\mathbb Z_2^n$ are induced by the above mentioned right action
$$
X \mapsto Xg, \quad P \mapsto Pg, \quad g\in SO(r,{\mathbb C}),
$$
which yields
\begin{gather}
{\bf L}(\la) = \begin{pmatrix} \qquad \la^{n-1} {\mathcal V}_0
+ \cdots &
\la^{n-1} {\bf I}_r  + \cdots \\
\la^{n} {\bf I}_r + \la^{n-1}{\mathcal W}_0 + \cdots &
- \la^{n-1} {\mathcal V}_0^T  + \cdots \end{pmatrix} \longmapsto \notag \\
 \begin{pmatrix} g^T & 0 \\ 0 & g^T \end{pmatrix} {\bf L}(\la) \begin{pmatrix} g & 0 \\ 0 & g \end{pmatrix}
 =\begin{pmatrix} \qquad \la^{n-1} g^T {\mathcal V}_0 g + \cdots &
\la^{n-1} {\bf I}_r  + \cdots \\
\la^{n} {\bf I}_r + \la^{n-1}g^T {\mathcal W}_0 \, g + \cdots & -
\la^{n-1} g^T {\mathcal V}_0^T g  + \cdots \end{pmatrix}, \label{action_g}
\end{gather}
The action changes neither the structure of ${\bf L}(\la)$, nor the
spectral curve, but may change the block ${\mathcal V}_0=\frac 12
\Psi$ of the $\mathrm{so}(r)$-momenta integrals. If, instead, $g$
belongs to the stabilizer of ${\mathcal V}_0$ (that is, $({\mathbb
C}^*)^{[r/2]}$), 
the latter is preserved as well.

A point of ${\cal I}_{\text {red}}= {\cal I} /SO(r,{\mathbb
C})/{{\mathbb Z}_2^n }$ defines an equivalence class $\{{\cal
D}\}$ of effective divisors on ${\cal S}'$ via the sequence of maps
\begin{equation} \label{EM}
(X,P) \longmapsto \mathbf L(\lambda) \longmapsto \psi \longmapsto {\cal D}
\end{equation}
 (where, as above, $\langle \alpha,\psi\rangle=1$), and we get the
injective eigenvector map\footnote{The injectivity of the map follows from the injectivity of eigenvector map on
${\mathcal I}_{S'}/\mathbb{PGL}(2r,{\mathbb C})$ described in the subsection \ref{EVM}.}
$$
{\cal M} \colon  {\cal I}_{\text{red}} \longrightarrow \Jac ({\mathcal S}') .
$$
Note that, in view of Proposition \ref{dims}, the dimensions of
$\mathcal I_{\mathrm{red}}$ and of $\Prym ( {\mathcal S}' , \sigma)\subset \Jac ({\mathcal S}')$
coincide. We now show that $\mathcal M$ maps the
invariant manifold $\mathcal I_{\mathrm{red}}$ to a translated
Prym subvariety.

\begin{prop} \label{sub_Prym} For any equivalence class $\{(X,P)\}$  in ${\cal I}_{red}$,
the divisor ${\cal D}={\cal M}(\{( X,P) \})$ satisfies the relation
\begin{equation} \label{z_p}
{\cal D} + \sigma {\cal D} \equiv {\cal B} ,
\end{equation}
where ${\cal B}$ is the set of all branch points of the projection ${\mathcal S}'\to {\mathbb P}\{ \la\}$,
including the infinite points $\infty_1,\dots, \infty_r$.
\end{prop}

Then, since ${\cal B}$ is independent of $(X,P)\in {\cal
I}_{red}$, the degree zero divisor ${\cal D} - N P_0$ with a fixed
basepoint $P_0\in {\cal S}'$ belongs to a translated Prym variety
$\Prym ( {\mathcal S}' , \sigma)$.
Since $\cal M$ is injective and ${\cal
I}_{red}$ and $\Prym ( {\mathcal S}' , \sigma)$ have the same
dimension $l$, we proved

\begin{prop} \label{red_H}
The reduced invariant manifold ${\cal I}_{\text {red}}= {\cal I} /SO(r,{\mathbb C})/{{\mathbb Z}_2^n}$ of the
Neumann system is isomorphic to an open subset $\Prym_0 ({\mathcal S}', \sigma)$ of $\Prym ( {\mathcal S}' , \sigma)$.
\end{prop}


\noindent{\it Proof of Proposition} \ref{sub_Prym}.
Here we use the technique already applied in \cite{Bog, MV, Dub_0}. Namely, let
$$
\psi(P)=  \left( \chi (P) , \xi (P) \right )^T,  \quad \chi,\xi \in {\mathbb C}^r, \quad P=(\la,w) \in {\cal S}'
$$
be an eigenvector of ${\bf L}(\la)$ with the eigenvalue $w$,
and  $\bar\psi(P)= \psi(\sigma P) = (\bar\chi, \bar\xi)^T$ be an eigenvector with the opposite eigenvalue:
${\bf L}(\la) \bar\psi(P) =- w \bar\psi(P)$.
Introduce also the vector $\psi^*(P) =(-\bar\xi,\bar\chi)^T$, which, due to the structure of ${\bf L}(\la)$ in
\eqref{symp_form}, is an eigenvector of ${\bf L}^T (\la)$ with the eigenvalue $w$:
\begin{equation} \label{eigen_transpose}
{\bf L}^T (\la) \psi^*(P) = w \, \psi^*(P).
\end{equation}
Indeed,
$$  {\bf L}^T \begin{pmatrix} -\bar\xi \\ \bar\chi \end{pmatrix}
 = \begin{pmatrix} {\cal V}^T & {\cal W} \\ {\cal U} & -{\cal V} \end{pmatrix}
\begin{pmatrix} -\bar\xi \\ \bar\chi \end{pmatrix} =
\begin{pmatrix} -{\cal V}^T \bar\xi + {\cal W}\bar\chi \\ -{\cal U} \xi -{\cal V}\bar\chi \end{pmatrix}
= \begin{pmatrix} -w \bar\xi \\ w \bar\chi \end{pmatrix} = w  \begin{pmatrix} - \bar\xi \\ \bar\chi \end{pmatrix} .
$$
Now consider the function
\begin{equation} \label{key_function}
F(P)=\langle \psi^*(P), \psi(P) \rangle = \psi^T (P) \psi^*(P) = \bar\chi \xi-\chi\bar\xi.
\end{equation}
The rest of the proof is based on item 1) of the following lemma that is proved in the Appendix.

\begin{lem} \label{Lem1}
\begin{description}
\item{1)}
The function $F(P)$ has simple zeros only at the branch points of $w$ as the function of $\la$ and has the poles
only at $\cal D$ and $\sigma {\cal D}$.
\item{2)}
$F(P)$ is anti-symmetric with respect to the involution $\sigma\, : {\cal S}'\to {\cal S}'$.
\end{description}
\end{lem}

Since $F(P)$ is meromorphic on ${\mathcal S}'$, the divisors of
its zeros and poles are equivalent, which leads to the relation
\eqref{z_p} and Proposition \ref{sub_Prym}.\hfill $\square$

\medskip

Since for an appropriate normalization of the eigenvector
$\psi(P)$ the pole divisor $\cal D$ is finite, the eigenvector map
${\cal M}\colon {\cal I}_{\text{red}} \longrightarrow  \Jac
({\mathcal S}')$ can be extended to the map
$$
\widetilde{\cal M}\colon {\cal I}_h /{{\mathbb Z}_2^n}
\longrightarrow  \widetilde\Jac ({\mathcal S}',\infty)
$$

\begin{prop}
The map $\widetilde{\cal M}$ is injective.
\end{prop}

\noindent{\it Proof.} The idea is borrowed from \cite{Gavr}.
Namely, let $\psi(P)$, $P\in \mathcal S'$ be the eigenvector
bundle of a Lax matrix ${\bf L}(\la)\in {\cal I}_h /{{\mathbb
Z}_2^n}$ with the normalization \eqref{norm}, and $\cal D$ be the
corresponding effective divisor on $\mathcal S'$ defining a point
${\cal D}-NP_0$ in $\widetilde \Jac( {\mathcal S}',\infty)$. We
will show that, given the $\infty$-equivalence class of $\cal D$,
the matrix ${\bf L}(\la)$ can be reconstructed uniquely.

Let $\tilde D$ be another effective divisor giving the same point
in $\widetilde \Jac( {\mathcal S}',\infty)$. Then, by the
generalized Abel theorem, there exists a meromorphic function
$G(P)$ on $S'$ such that
$$
  (G)= {\cal D}- \tilde{\cal D}, \qquad G(\infty_s^-)=G(\infty_s^+)\ne 0,\infty, \quad s=1,\dots,[r/2].
$$
The components of $\psi(P)$ form a basis of $L({\cal
D})$\footnote{Note that these components, in general, take
different values at $\infty_s^-$ and $\infty_s^+$.}. Respectively,
let $\tilde{\psi}(P)$ be a vector bundle whose components form a
basis of $L(\tilde{\cal D})$.

Then the components of $G(P) \tilde \psi(P)$ form a basis of
$L({\cal D})$, and there is a non-degenerate matrix $R\in
{GL}(2r,{\mathbb C})$ independent of $P$ such that $\psi (P)= G(P)
R \, \tilde \psi(P)$. The bundles $\psi (P), \tilde \psi(P)$
define the same matrix ${\bf L}(\la)$ up to conjugation by $R$. By
\eqref{action_g}, the conjugation preserves ${\bf L}(\la)\in {\cal
I}_h/{{\mathbb Z}_2^n } $  if and only if
$R=\diag(g,g)$,
where $g\in SO(r,{\mathbb C})$
 is a stabilizer of ${\cal V}_0 \in \mathrm{so}(r, {\mathbb C})$.

On the other hand, the above relation between $\psi, \tilde\psi$
implies
$$
\psi(\infty_k) = G(\infty_k) R \, \tilde\psi(\infty_k) , \qquad
k=1,\dots, r .
$$
Due to the chosen normalization, near $\infty_k$ the components of
$\psi(\infty_k), \tilde\psi(\infty_k)$ are finite and given by the
expansions \eqref{exp_psi} involving the eigenvectors ${\bf v}_s,
\bar{\bf v}_s, {\bf v}_0$ of ${\cal V}_0$. This implies
$$
   {\bf v}_s || \, G(\infty_s) g \,{\bf v}_s , \quad  \bar{\bf v}_s || \, G(\infty_s) g \,\bar{\bf v}_s, \qquad
 s=1,\dots, [r/2], \quad {\bf v}_0 || \, G(\infty_r) g \,{\bf v}_0
$$
hence all ${\bf v}_s, \bar{\bf v}_s, {\bf v}_0$ must be also
eigenvectors of $g$. Since $g\in SO(r,{\mathbb C})$, we conclude
that $g$ and $R$ are the unit matrices, therefore the divisors
${\cal D}, \tilde{\cal D}$ correspond to the same matrix ${\bf
L}(\la)\in {\cal I}_h/{{\mathbb Z}_2^n}$. \hfill$\square$

\begin{prop} \label{PRYM}
The image $\widetilde {\cal M}({\cal I}_h/\mathbb Z_2^n)$ belongs to a translate of
$\widetilde\Prym ({\mathcal S}', \sigma)$ in $\widetilde{\Jac}({\mathcal S}')$.
\end{prop}

\noindent{\it Proof.} The zeros and poles of the meromorphic function $F(P)$ given by \eqref{key_function} imply the equivalence
\eqref{z_p} on $\Jac({\cal S}')$, namely ${\cal D} + \sigma {\cal D} \equiv {\cal B}$,
 which cannot be extended to the equivalence
${\cal D} + \sigma {\cal D} \equiv_\infty {\cal B}$,
because $ F(\infty_s^{\pm})=0$ and the divisor ${\cal B}$ contains all the infinite
points:
$$
{\cal B}={\cal B}_0 + \infty_1 + \cdots + \infty_r \, ,
$$
${\cal B}_0$ being the finite part of $\cal B$.  
This can be repaired by introducing the function
$G(P)=F^2(P) \lambda$, where $\lambda$ is the coordinate on the curve ${\mathcal S}'\subset {\mathbb P}^2 (\lambda,w)$ and
$$
(\lambda)= {\cal O} - 2(\infty_1 + \cdots + \infty_r) ,
$$
$\cal O$ being the preimage of $\lambda=0$ on ${\mathcal S}'$. Then
$$
   (G) = 2 {\cal B}- 2({\cal D} +\sigma {\cal D}) + {\cal O}- 2(\infty_1+\cdots+\infty_r) =
2{\cal B}_0 + {\cal O}- 2({\cal D} +\sigma {\cal D}) .
$$
In view of Lemma \ref{Lem1} and the symmetry $\lambda(\sigma P)=\lambda(P)$,
$$
  G(\infty_s^-)= G(\infty_s^+) \ne 0, \infty , \qquad s=1,\dots,[r/2] .
$$
Hence the expression for $(G)$ yields
 \begin{equation} \label{eq_gen}
  2( {\cal D} +\sigma {\cal D} ) \equiv_\infty 2{\cal B}_0 + {\cal O},
\end{equation}
The latter implies that ${\cal D} +\sigma {\cal D}$ is a fixed divisor defined up to 
translations by half-periods $\cal P$ of $\widetilde\Jac(\mathcal S',\infty)$.
Then, for a certain translate $\Sigma$ of $\widetilde\Prym(\mathcal S',\sigma)\subset
\widetilde\Jac({\cal S}',\infty)$, the divisor  ${\cal D}-N P_0$ belongs to a union of translates of $\Sigma$
by the half-periods.  

On the other hand, comparing \eqref{eq_gen} with \eqref{z_p}, one
sees that any half-period $\mathcal P$ must satisfy $\phi(\mathcal P)=0$, where $\phi$ is specified in \eqref{ex_sec}.
Therefore, $\mathcal P$ includes only the half-periods $\upsilon(\pm 1,\dots, \pm 1)$,
which, according to Lemma \ref{lem_ups}, are half-periods in
$\widetilde\Prym(S',\sigma)$. Hence, the translates of $\Sigma$ by
$\mathcal P$ coincide with $\Sigma$, which proves the proposition.
\hfill$\square$
\medskip

Since ${\mathcal I}_h$ and $\widetilde\Prym(\mathcal S',\sigma)$
have the same dimension $\delta=l+[r/2]$ and the extended eigenvector map $\widetilde{\cal M}\colon {\cal I}_h/{{\mathbb Z}_2^n}  \longrightarrow  \widetilde\Jac
({\mathcal S}',\infty)$ is injective, we arrive at the main
theorem of the section.

\begin{thm}\label{MAIN}
If the rank of the momentum $\Psi$ is maximal, then, after factorization by the reflection group ${\mathbb Z}_2^n$,
a generic $\delta$-dimensional complex invariant manifold ${\mathcal I}_h$ of the Neumann system on
$T^*_{\mathbb C} V_{n,r}$
is $\widetilde\Prym_0 ({\mathcal S}', \sigma)$, an open subset of a translate of the affine Prym variety
$\widetilde\Prym ({\mathcal S}', \sigma)\subset \widetilde{\Jac}({\mathcal S}', \infty)$.
\end{thm}

This theorem together with Proposition \ref{red_H} can be summarized with the following commutative diagram
\begin{equation} \label{cd_fin}
\begin{CD}
{\cal I}_h/{{\mathbb Z}_2^n }  @> \widetilde{\cal M}  >> \widetilde\Prym_0 ({\mathcal S}', \sigma) \\
@ V ({\mathbb C}^*)^{[r/2]} VV @ VV \phi V \\
{\cal I}_{\text{red}}  @ > {\cal M} >> \Prym_0 ({\mathcal S}',
\sigma)
\end{CD}
\end{equation}


\subsection{Linearization of the flow}
We now show that under
the eigenvector maps $\widetilde{\cal M}\colon {\cal I}_h /{{\mathbb Z}_2^n} \longrightarrow \widetilde\Jac
({\mathcal S}',\infty)$ and ${\cal M} \colon  {\cal I}_{\text{red}} \longrightarrow \Jac ({\mathcal S}')$,
 the trajectories $(X(t),P(t))$ of the complex Neumann system
induce straight line trajectories on the (generalized) Jacobian variety.

Let, as above, $\psi(P,t)$ be eigenvector of $\mathbf L(\lambda)=a(\lambda)L(\lambda)$
with a normalization $\langle\alpha,\psi(P)\rangle =1$. 
As follows from the Lax representation \eqref{LA3},
$\dot\psi+N_\kappa \psi$ is also an eigenvector of $\mathbf L(\lambda)$. Therefore, it is proportional to
$\psi(P)$ with a meromorphic multiplier $f$:
$$
\dot\psi(P)+N_\kappa(\la) \psi(P)=f (P,t) \psi(P).
$$
Since $\frac{d}{dt} \langle\alpha,\psi(P)\rangle =0$, the above implies
$f(P,t)=\langle\alpha,N_\kappa\psi\rangle$.

Now assume that $\psi(P)$ is normalized as in Proposition \ref{puis}, so its behavior near the infinite
points $\infty_1,\dots, \infty_r$ is given by the expansions \eqref{exp_psi}.
Then, near $\infty_s$ with the local coordinate
$\tau$ ($\lambda=1/\tau^2$), the following expansions hold
\begin{align}
\nonumber N_\kappa\psi =& \begin{pmatrix} \kappa\,\mathcal V_0 & \mathbf{I}_r \\
\tau^{-2} \mathbf{I}_r + \Lambda & \kappa \mathcal V_0
 \end{pmatrix}
\begin{pmatrix} {\bf v}_s \tau + {\bf h}_s \tau^2+O(\tau^3) \\
{\bf v}_s  + {\bf h}_s\tau +O(\tau^2 ) \end{pmatrix} \\
 =& \begin{pmatrix} {\bf v}_s + (\kappa\nu_s {\bf v}_s+\mathbf h_s)\tau_s+ O(\tau^2) \\
 \tau^{-1} {\bf v}_s +\kappa\nu_s{\bf v}_s + \mathbf h_s + O(\tau)
 \end{pmatrix},   \label{expansion1}
\end{align}
where for $j=1,\dots,[r/2]$, $\mathbf v_{j+[r/2]}=\bar{\mathbf v}_j$,
$\mathbf h_{j+[r/2]}=\bar{\mathbf h}_j$, $\nu_{[r/2]+j}=-\nu_j$,
and, if $r$ is odd, $\mathbf v_r=\mathbf v_0$, $\mathbf h_r=\mathbf h_0$, $\nu_r=0$.
The chosen normalization \eqref{norm} implies
\begin{gather*}
\langle\ \beta, {\mathbf v}_s + {\mathbf h}_s \tau +O(\tau^2) \rangle \equiv 1, \\
f(\tau) = \langle \alpha, N_\kappa\psi\rangle =
\langle \beta, \tau^{-1}[ {\mathbf v}_s + {\mathbf h}_s \tau]
+\kappa \nu_s {\mathbf v}_s +O(\tau) \rangle .
\end{gather*}
As a result, near $\infty_1,\dots, \infty_r$, the function $f(P,t)$ has the expansions
\begin{align}
\nonumber\infty_s^+\colon\; &
f = \frac{1}{\tau} + \kappa \nu_s + O(\tau),  \\
\infty_s^-\colon\;& \label{exp-f}
f= \frac{1}{\tau} - \kappa \nu_s + O(\tau), \quad s=1,\dots,[r/2],\\
\nonumber\infty_0\colon\;&
f = \frac{1}{\tau} + O(\tau)
\quad\qquad\qquad\qquad (\text{if {\it r} is odd}),
\end{align}
regardless to the choice of $\beta$. 

Let $\mathcal D_t$ be the divisor of poles of $\psi(P,t)$ and
$\tilde{\mathcal A}(\mathcal D_t)$ be its image under the generalized Abel map \eqref{GAM} with holomorphic differentials
$\omega_1,\dots,\omega_g$ and meromorphic differentials of the 3rd kind $\varOmega_j$, $j=1,\dots,[r/2]$.
Assume now that the latter are {\it normalized}: they
have simple poles with the residua $\pm 1$ at $\infty_j^+, \infty_j^-$ respectively, and no poles elsewhere.

\begin{thm}\label{linear2}
For generic values of the first integrals,
the map $\widetilde{\mathcal M}$ linearizes the complex Neumann
flow on  the generalized Jacobian $\widetilde{\Jac}({\mathcal S}', \infty)$ as follows
\begin{align}
\nonumber\tilde{\mathcal A}(\mathcal D_t)
-\tilde{\mathcal A}(\mathcal D_0)=&-t\sum_{k=1}^r \mathrm{Res}_{\infty_k}
(f\omega_1,\dots,f\omega_g, f\varOmega_{1},\dots,
f\varOmega_{[r/2]} )^T \\
&-t(0,\dots,0,2\kappa\nu_1,\dots,2\kappa\nu_{[r/2]})\quad\qquad
(\mathrm{mod}\,\widetilde\varLambda).
\label{new-equation}
\end{align}
\end{thm}

In particular, under the customary eigenvector map $\mathcal M$, the complex Neumann flow gives the
following flow on $\Jac({\cal S}')$
\begin{equation}\label{new-equation2}
\mathcal A(\mathcal D_t)-\mathcal A(\mathcal D_0)=-t\sum_{s=1}^r
\mathrm{Res}_{\infty_s} (f\omega_1,\dots,f\omega_g)^T\qquad
(\mathrm{mod}\,\varLambda).
\end{equation}

Observe that as the residua of $f$ at $\infty_k$ do not depend on $t$,
the flows \eqref{new-equation}, \eqref{new-equation2} are indeed linear ones.

Note also that in the classical case $r=1$ (the Neumann system on $S^{n-1}$ linearized on the Jacobian of the
hyperelliptic curve $\Gamma$)
\eqref{new-equation2} gives a correct direction vector on $\Jac (\Gamma)$, whose
components are the leading terms of the expansion of $g=n-1$
holomorphic differentials near the infinite point of $\Gamma$ (see e.g., \cite{Mum}).

The proof of Theorem \ref{linear2} is an extension of that of
Theorem 6.39 in \cite{AMV} describing linearization of equations
admitting polynomial Lax representations, and it is given in Appendix.

The direction of the flow \eqref{new-equation} can be written more specifically
if we choose a basis of $H^1(\mathcal S',{\mathbb Z})$ in the form
$(\omega_1^- ,\dots,\omega_l^-, \omega_1^+, \dots, \omega_{g-l}^+)$,
where $\omega_i^-$, $\omega_j^+$ are
anti-symmetric and symmetric holomorphic differentials
respectively, and, as above, $l$ is the dimension of $\Prym ({\mathcal S}', \sigma)$.
 Since the involution $\sigma$ flips sign of the local parameter $\tau$ (see \eqref{sigma3}, \eqref{sigma2}),
near the infinite points the differentials admit expansions
\begin{align*}
\infty_s^+\colon & \quad \omega_i^- =( \chi_{i,s}+ O(\tau))d\tau, \quad
\omega_j^+ =(\xi_{j,s} + O(\tau))d\tau, \\
\infty_s^-\colon & \quad \omega_i^- =(\chi_{i,s}+ O(\tau) )d\tau,  \quad
\omega_j^+ =-(\xi_{j,s} + O(\tau))d\tau,  \\
\infty_0 \colon  & \quad  \omega_i^- =( \chi_{i,r}+ O(\tau))d\tau, \quad
\omega_j^+ =O(\tau)d\tau,
\end{align*}
where $\chi_{i,s}, \xi_{j,s}$, $s=1,\dots,[r/2]$, $i=1,\dots,l$, $j=1,\dots,n-l$ are some constants.
Also, choose the meromorphic differentials satisfying
$\sigma^* \varOmega_{{k}} = - \varOmega_{{k}}$.
Again, according to \eqref{sigma3}, \eqref{sigma2}, near $\infty_s^+, \infty_s^-$ we have
\begin{equation}
\label{exp_Om}
\varOmega_{k}= \Big( \frac {\delta_{ks}}{\tau_s} + \phi_{k,s} +
O(\tau) \Big) d\tau, \quad \varOmega_{k}
= \Big(- \frac {\delta_{ks}}{\tau} + \phi_{k,s} + O(\tau) \Big) d\tau,
\end{equation}
where $\delta_{ks}$ is the Kronecker symbol and $\phi_{k,s}$, $k,s=1,\dots,[r/2]$ are some constants.

Theorem \ref{linear2} implies that in the above basis of differentials
the direction of the flow on $\widetilde\Jac({\cal S}', \infty)$ is
\begin{gather}
\label{direction2}
-\big(\chi_1\, ,\, \dots\, ,\, \chi_l\, ,\, \underbrace{0\, ,\,
\dots\, ,\, 0}_{g-l \; \text{times}}\, ,\,  4\kappa \,\nu_1 +
\phi_1\, , \,\dots \,,\,  4\kappa\, \nu_{[r/2]} + \phi_{[r/2]} \big)^T\, , \\
 \chi_i=\chi_{i,r}+ 2 \sum_{s=1}^{[r/2]} \chi_{i,s}, \qquad \phi_k = 2\sum_{s=1}^{[r/2]} \phi_{k,s}. \notag
\end{gather}

Consequently, the direction of the reduced flow on $\Jac({\cal S}')$ is
\begin{equation}\label{direction1}
-\big(\chi_1\, ,\, \dots\, ,\, \chi_l\, ,\, \underbrace{0\, ,\,
\dots\, ,\, 0}_{g-l \; \text{times}}  \big)^T\, ,
\end{equation}
which shows that the flow goes along $\Prym ( {\mathcal S}' ,
\sigma)\subset \Jac({\cal S}')$, as expected.

\begin{rem}{\rm The reduced Neumann flow on $T^*V_{n,r}/SO(r)$
 is described by the flow \eqref{new-equation2} on $\Prym ({\mathcal S}' , \sigma)$
whose direction \eqref{direction1} does not depend on the
parameter $\kappa$ in the metric on $V_{n,r}$. This agrees with
the fact that the metric term $\kappa\langle\Psi,\Psi\rangle$
leads to a trivial vector field on $T^*V_{n,r}/SO(r)$. On the contrary,
the original, non-reduced problem on $T^*V_{n,r}$ is
linearized on $\widetilde\Prym ({\mathcal S}', \sigma)$, and the
direction of the flow \eqref{direction2} depends on $\kappa$.
}\end{rem}

\paragraph{Linearization of the Neumann system on the Grassmannian variety.}
The Neumann system on the Grassmmanian $G_{n,r}$ is completely integrable
in the Liouville sense.
Let $\mathcal I_{G_{n,r}}\subset T^*_\mathbb C G_{n,r}$ be a generic, $r(n-r)$--dimensional complex level set of commuting integrals.
Since the Neumann system on $G_{n,r}$ can be seen as the $SO(r)$--reduction of a Neumann system on $V_{n,r}$ for the zero value
of the momentum mapping $\Psi$, we have the  natural identification:
$$
\mathcal I_{G_{n,r}}/\mathbb Z_2^n=
{\cal I}_{red}={\cal I}/SO(r,{\mathbb C})/{{\mathbb
Z}_2^n},
$$
where $\mathcal I=\mathcal I_0$ is the $(r(n-r)+r(r-1)/2)$--dimensional invariant manifold of the Neumann flow on the Stiefel variety
in the special case $\Psi=0$.
From Theorem \ref{big_g} and item 2) of Proposition \ref{dims}, we have
$$
\mathrm{gen}(\mathcal S')=\dim\Jac({\cal S}')=2r(n-r) -n+1, \quad \dim\Prym({\mathcal S}', \sigma)=r(n-r).
$$
Besides, like in the general case,
Proposition \ref{sub_Prym} implies that the map $\cal M$ realized a bijection
between ${\cal I}_{red}$ and an open subset of $\Prym({\mathcal S}', \sigma)$.

The above considerations can be summarized in the following statement.

\begin{thm}\label{LinGr} For generic values of the first integrals of the complex Neumann system on the Grassmannian,
the eigenvector map
$$
{\mathcal M}\colon   \mathcal I_{G_{n,r}}/\mathbb Z_2^n \longrightarrow \Jac({\cal S}')
$$
linearizes the flow on an open subset of a translate of the Prym variety $\Prym({\mathcal S}', \sigma)$.
\end{thm}

\begin{rem}{\rm
It can be proved that generic trajectories
of the Neumann flows on the unreduced space with the zero momentum $\Psi$ filled up tori of dimension $r(n-r)$, while
the level sets of the integrals
$\{f_{k,l},\Psi_{ij}\}$ have the dimension $r(n-r)+r(r-1)/2$.
That is, over each
torus in the factor $\Psi^{-1}(0)/SO(r)$ there is
$SO(r)$-parametric family of tori in $\Psi^{-1}(0)$.
On the algebraic-geometric side this can be seen as follows.
In the special case $\Psi=0$ all $r$
infinite points of the regularized curve ${\cal S}'$ are invariant
with respect to the involution $\sigma$ (see \eqref{sigma4}). Such a situation has been
studied in \cite{be77}: one considers the generalized
Jacobian $\widetilde{\Jac}({\mathcal S}', \infty)$ of the
singularized curve, obtained from ${\cal S}'$ by gluing
$\infty_1,\dots, \infty_r$ to one point. In contrast to the
general case, now the part of $\widetilde{\Jac}({\mathcal S}',
\infty)$, which is anti-invariant with respect to $\sigma$, is
just the compact subvariety $\Prym ({\mathcal S}', \sigma)$. Then the corresponding extended eigenvector map
$\widetilde{\mathcal M}\colon \mathcal I_0/\mathbb Z^2_n \to \widetilde{\Jac}({\mathcal S}',
\infty)$ linearise a complex trajectory over a translate of $\Prym ({\mathcal S}', \sigma)$.
}
\end{rem}

\section{Integrable discretization on $T^*V_{n,r}$} \label{DDis}

\subsection{The map and its basic properties}

Using the notation of  Section \ref{NSV}, consider the
following matrix generalization of the map (\ref{moms}) (discrete Neumann system) with a real parameter $\la_*$
\begin{equation} \label{moms_r}
\begin{aligned}
P & =\;  A^{1/2}(\la_*)\, \tilde X - X\, \Gamma(\la_*), \\
\tilde P & = - A^{1/2}(\la_*)\, X + \tilde X\, \Gamma(\la_*) ,
\end{aligned}
\end{equation}
where, as above, $A(\la)=\la {\bf I}_n -A$ and $\Gamma$ is a {\it
symmetric} $r\times r$ matrix, which is found from the condition
$X^T P+P^T X=0$:
\begin{equation} \label{gam}
\Gamma (\la_*) = \frac 12 \big( \tilde X^T A^{1/2}(\la_*)X + X^T A^{1/2}(\la_*)\tilde X \big).
\end{equation}

The first matrix equation in (\ref{moms_r}) gives the definition of the discrete momentum $P$,
whereas the second equation provides the
discrete dynamics in an implicit form. In view of (\ref{gam}), the structure of (\ref{moms_r}) ensures preservation of the $SO(r)$ momentum:
$\tilde X^T \tilde P= X^T P$.

Equations (\ref{moms_r}), (\ref{gam}) define a a multi-valued map (a correspondence)
$$
{\mathfrak B}_r\, :\, T^*V_{n,r} \longrightarrow T^* V_{n,r} ,
\qquad {\mathfrak B}_r (X,P)=(\tilde X, \tilde P),
$$
which can be regarded as a discrete Neumann system on $T^* V_{n,r}$\footnote{
In fact, it can be considered without the restriction $X^TP+ P^T X=0$, i.e., can be extended
to the map on $V_{n,r} (X)\times \R^{nr}(P)$, which always preserves the matrix $X^TP$.}.

To evaluate the map, we rewrite (\ref{moms_r}) in the form
\begin{equation} \label{moms_r*}
\begin{aligned}
\tilde X & =\, A^{-1/2}(\la_*)(P+ X\Gamma(\la_*)), \\
\tilde P & = - A^{1/2}(\la_*)X + A^{-1/2}(\la_*)\,(P+ X\Gamma(\la_*))\, \Gamma(\la_*) .
\end{aligned}
\end{equation}

Then, applying the condition $\tilde X^T \tilde X = {\bf I}_r$, we
obtain the following quadratic matrix equation for $\Gamma$, which
defines it implicitly as a function of $X,P$ and which generalizes
the scalar equation (\ref{quad_beta}),
\begin{equation}
\label{quad_gamma}
\Gamma  {\bf U} \Gamma + \Gamma {\bf V}+ {\bf V}^T \Gamma - {\bf
W}=0,
\end{equation}
where
\begin{equation}
{\bf U}= X^T A^{-1}(\la_*) X , \quad  {\bf V}= X^T A^{-1}(\la_*)
P, \quad {\bf W}= {\bf I}_r - P^T A^{-1}(\la_*)P. \label{blocks}
\end{equation}

This is a system of $r(r+1)/2$ scalar quadratic equations for the
$r(r+1)/2$ components of $\Gamma$. However, due to the structure
of the matrix coefficients, the number of its solutions is less
than that predicted by the Besout theorem. The matrix equation
(\ref{quad_gamma}) is known in the literature in connection with
stationary solutions of the matrix Riccati differential equation,
optimum automatic control theory, and its complete solution was
presented by Potter \cite{Pot}.

In order to describe it, we first observe that the
coefficients (\ref{blocks}) coincide with the $r\times r$ blocks
of the Lax matrix $L(\la_*)$ in \eqref{Lax_r}. Since the latter is
symplectic, its eigenvalues $w_1,\dots,w_{2r}$ are divided into
$r$ pairs $(w_i,-w_i)$. Let $\psi_1,\dots, \psi_r\in {\mathbb
C}^{2r}$ be the eigenvectors of $L(\la_*)$ with {\it distinct}
eigenvalues $w_1,\dots,w_r$ such that $w_i\ne -w_j$. The
corresponding matrix $\Psi=(\psi_1\, \cdots \,\psi_r)$ will be
called a {\it non-special eigenmatrix}.

The following proposition is a direct consequence of the results
of \cite{Pot}.

\begin{prop} \label{quad_sol} Any symmetric solution of the matrix quadratic equation \eqref{quad_gamma} has the form
$$
\Gamma = \Upsilon \, \Xi^{-1},
$$
where $\Xi, \Upsilon$ are upper and lower $r\times r$ halves of a non-special eigenmatrix
$\Psi = \begin{pmatrix} \Xi \\ \Upsilon \end{pmatrix}$ of $L(\la_*)$.
\end{prop}

\begin{rem}\label{POTER}{\rm
The original paper \cite{Pot} solves the equation
\begin{equation}\label{pot-eq}
 \Gamma  {\mathfrak U} \Gamma + \Gamma{\mathfrak V}+ \bar{\mathfrak V}^T \Gamma - {\mathfrak W}=0,
\end{equation}
where $\mathfrak U, \mathfrak V, \mathfrak W$ are arbitrary
$n\times n$ complex matrices. As above, consider the matrix
$$
K=
\begin{pmatrix} \mathfrak V & \mathfrak U \\ \mathfrak W & -\bar{\mathfrak V}^T
\end{pmatrix}.
$$
According to \cite{Pot}, if $K$ has a diagonal Jordan canonical
form with {distinct} eigenvalues $w_1,\dots,w_r$ such that $w_i\ne
-\bar w_j$, then the {\it hermitian solutions}
($\bar\Gamma^T=\Gamma$) of \eqref{pot-eq} are exactly those
described in Proposition \ref{quad_sol}: $\Gamma = \Upsilon \,
\Xi^{-1}$, where $\Xi, \Upsilon $ are upper and lower $r\times r$
halves of the corresponding eigenmatrix of $K$. Proposition
\ref{quad_sol} dealing with symmetric matrices can be easily
proved following the lines of \cite{Pot}.
}\end{rem}

Obviously, the product $\Upsilon  \, \Xi^{-1}$ is invariant under
any gouge transformation $\Psi \mapsto \Psi\, G$, $G\in GL(r,C)$.
As was also shown in \cite{Pot}, if $\Psi$ is special (that is,
contains eigenvectors $\psi_i, \psi_j$ with $w_i=-w_j$) or its
columns are linear combinations of more than $r$ eigenvectors
$\psi_i$, then the product $\Upsilon \, \Xi^{-1}$ also satisfies
(\ref{quad_gamma}), but it is {\it not} a symmetric matrix.

Since for generic finite $\la_*\ne a_i$ there are $2^r$ possible partitions
$$
\mathbf w=\{w_1,\dots,w_r \mid -w_1,\dots,-w_r\},
$$
 we conclude that the
equation (\ref{quad_gamma}) has precisely $2^r$ complex solutions.
They admit a natural decomposition $\{\Gamma \} = \{\Gamma_- \}
\cap \{\Gamma_+ \}$ such that for any solution $\Gamma^*\in
\{\Gamma_- \}$ corresponding to a partition $\mathbf
w_-=\{w_1,\dots,w_r \mid -w_1,\dots,-w_r\}$ there is a unique
$\Gamma^{**}\in \{\Gamma_+\}$ corresponding to the ''opposite''
partition $\mathbf w_+=\{-w_1,\dots,-w_r \mid w_1,\dots,w_r\}$.

Since the spectral curve ${\cal S}'$ has ordinary branch points $(\la=a_i,w=0)$,
in the special case $\la_* = a_i$
there are only $r-1$ distinct non-zero eigenvalues $w_1,\dots,w_{r-1}$ such that $w_i\ne -w_j$, and the number of
symmetric solutions of (\ref{quad_gamma}) drops to $2^{r-1}$.

Once a solution for $\Gamma$ is chosen, $\tilde X, \tilde P$ are
found uniquely from (\ref{moms_r}) or (\ref{moms_r*}). Since, for
generic $\la_*$, different solutions $\Gamma$ lead to different
images $\tilde X, \tilde P$, we conclude that the {\it complex}
map ${\mathfrak B}_r$ is $2^r$-valued.

\paragraph{Continuous limit.} Like in the case $r=1$,
the continuous limit of the map ${\mathfrak B}_r$ is obtained by
letting $\la_*\to\infty$. Namely, we set $\la_*=1/\epsilon^2$
and consider the expansions
\begin{gather*}
A^{-1}(\la_*)= t^2 \left( {\bf I}_n + {\epsilon^2} A +O(\epsilon^4 ) \right) , \\
\tilde X= X+ \epsilon \dot X+ O(\epsilon^2), \quad \tilde P = P+ \epsilon \dot
P+ O(\epsilon^2) .
\end{gather*}
This gives
\begin{equation} \label{expan_G}
\Gamma = \frac 1 \epsilon {\bf I}_r - \frac \epsilon 2 \big( X^T A X + P^T P\big) +
O(\epsilon^2 ) .
\end{equation}
Indeed, substituting the above into the matrix quadratic equation
\eqref{quad_gamma} and using the expansions
\begin{align*}
&{\bf U}  = \epsilon^2 \big( {\bf I}_r + \epsilon^2 X^T A X+ O(\epsilon^3) \big), \\
&{\bf V}  = \epsilon^2 \big( X^TP + \epsilon^2 X^T A P + O(\epsilon^3)  \big), \\
&{\bf W}  = {\bf I}_r - \epsilon^2 \big(P^T P  + \epsilon^2 P^T A P+
O(\epsilon^3)\big),
\end{align*}
we find that the coefficients at $\epsilon^0, \epsilon, \epsilon^2$ vanish,
which justifies \eqref{expan_G}.

Next, substituting the latter into the equations (\ref{moms_r*}),
and using
$$
A^{-1/2}(\la_*)= \epsilon \big( {\bf I}_n +\frac {\epsilon^2} 2 A
+O(\epsilon^4 ) \big) , \quad A^{1/2}(\la_*)= \frac 1\epsilon \big(
{\bf I}_n - \frac {\epsilon^2} 2 A + O(\epsilon^4) \big),
$$
we conclude that $\dot X, \dot P$ coincide with the right hand
sides of the continuous Neumann system with the Euclidean metric
\eqref{No}, \eqref{multiplier}, $\kappa=-1/2$.

\paragraph{Lagrangian description.}
The discrete Neumann system (\ref{moms_r}) can be considered as a
discrete variational problem on the Stiefel variety as well.
Namely, let $\{(X_k,P_k), k\in{\mathbb Z}\}$ be a trajectory of the
discrete Neumann system and $\{\Gamma_k\}$ be the corresponding sequence of matrix multipliers.
Comparing the first and the second equation in (\ref{moms_r}) with $k$
replaced by $k-1$, one obtains the equations
\begin{equation} \label{matr_quad}
X_{k-1}+ X_{k+1} = A^{-1/2}(\la_*) X_k B_k,  \qquad B_k = \Gamma_{k-1} + \Gamma_k .
\end{equation}
On the other hand, from \eqref{matr_quad} and the condition $X_{k+1}^TX_{k+1}=\mathbf I_r$
we get the matrix equation
$$
B_k(X_k^T A^{-1}(\la_*)X_k) B_k
-(X_{k-1}^TA^{-1/2}(\la_*)X_k)\,B_k - B_k\,(X_k^T
A^{-1/2}(\la_*)X_{k-1})=0,
$$
which determines $B_k$ as a function of $X_{k-1}$ and $X_k$ only.

Equations \eqref{matr_quad} coincides with the discrete
Euler-Lagrange equations on $V_{n,r}$
\begin{equation}\label{discrete-euler-lagrange}
\frac{\partial \mathcal L(X_k,X_{k+1})}{\partial X_k}+
\frac{\partial \mathcal L(X_{k-1},X_{k})}{\partial X_k}=X_k B_k,
\qquad k\in\mathbb Z
\end{equation}
of the functional $S=\sum_{k\in\mathbb Z} \mathcal L(X_k,X_{k+1})$ with the Lagrangian
\begin{equation}
\label{lagrangian} \mathcal L: V_{n,r} \times V_{n,r} \to\R, \quad
\mathcal L(X,\tilde X)= \tr(X^T A^{1/2}(\la_*) \tilde X),
\end{equation}
first derived by Moser and Veselov in \cite{MV}. In this sense, the expression for the
momentum $P$ in the system \eqref{moms_r} is actually the {\it
discrete Legendre transformation}:
\begin{equation}\label{DLegendreT}
\mathbf K: V_{n,r} \times V_{n,r} (X,\tilde X) \to T^*V_{n,r}(X,P) , \quad
P=\frac{\partial \mathcal L(X,\tilde X)}{\partial
X}-X\Gamma,
\end{equation}
where, as above, the multiplier $\Gamma$ is given by \eqref{gam}.
In particular, the correspondence $\mathfrak B_r$ is
symplectic (see \cite{MV, Ves2}).

\paragraph{The Lax representation.}
As we have seen above, the solutions of the matrix quadratic
equation (\ref{quad_gamma}) are closely related to the Lax matrix
of the continuous Neumann systems on $V_{n,r}$. It appears that
this matrix also forms a Lax representation of the discrete
system.

\begin{thm} \label{disr_Lax_Neum} Up to the action of the
group ${\mathbb Z}_2^n$ of reflections \eqref{reflection}, the
discrete Neumann system \eqref{moms_r}, \eqref{gam} is equivalent
to the intertwining matrix relation (discrete Lax pair)
\begin{equation} \label{disr_Lax_N}
\tilde L(\la) M (\la, \la_*)= M (\la, \la_*) L(\la),
\end{equation}
where
\begin{eqnarray*}
&& L (\lambda)=\begin{pmatrix}
X^T( \lambda \mathbf{I}_n-A)^{-1} P & X^T(\lambda \mathbf{I}_n-A)^{-1} X \\
\mathbf I_r - P^T(\lambda \mathbf{I}_n- A)^{-1} P & - P^T(\lambda
\mathbf{I}_n-A)^{-1} X
\end{pmatrix} , \notag \\
&& M (\la, \la_*) = \begin{pmatrix} - \Gamma (\la_*) & {\bf I}_r \\
 (\la - \la_*) {\bf I}_r +  \Gamma^2 (\la_*) &  - \Gamma (\la_*) \end{pmatrix}, \notag
\end{eqnarray*}
where $\tilde L(\la)$ depends on $\tilde X, \tilde P$ in the same
way as $L(\la)$ depends on $X, P$ and $M (\la, \la_*)$ depends on
$X, \tilde X$ in a symmetric way via \eqref{gam}.
\end{thm}

The proof is a direct computation (although quite a long one), it
uses the constraints (\ref{cond_XP}), the matrix identity
\begin{equation}\label{matrix-identity}
A(\lambda \mathbf{I}_n-A)^{-1}= (\lambda
\mathbf{I}_n-A)^{-1}A=\lambda(\lambda \mathbf{I}_n-A)^{-1}
-\mathbf{I}_n\, ,
\end{equation}
 and $SO(r)$--momentum preservation $\tilde
X^T \tilde P= X^T P$.

Note that the $2r\times 2r$ matrix $M$ is a direct generalization
of (\ref{MKuz}).

It follows that, regardless to the branch of the map ${\mathfrak B}_r$,
it preserves all the first integrals of the continuous
Neumann systems on $V_{n,r}$, including the non-commutative set given by the components of the $SO(r)$-momentum $\Psi$.
Thus, when the rank of $\Psi$ is maximal, generic invariant manifolds of ${\mathfrak B}_r$ are $l$--dimensional
isotropic tori.


\begin{thm}\label{NDI}
The discrete
Neumann system \eqref{moms_r} is completely integrable in the
noncommutative sense with the set of integrals given
by \eqref{perelomov} and by the components of the $SO(r)$-momentum
mapping $\Psi_{ij}$.
\end{thm}

The description of noncommutatively integrable symplectic correpondence is the same as that of a Liouville integrable
correspondence (see \cite{Ves2}), with the only difference that the Lagrangian tori
are replaced by the isotropic ones (see \cite{JJ}).

\paragraph{Double iterations.}  Although the map  ${\mathfrak B}_r$ is multi-valued, it has the following
return property, which generalizes that mentioned in Remark \ref{shift-rem}.

\begin{thm} \label{reiteration} Let $ {\mathfrak B}_r^-, {\mathfrak B}_r^+ $ be the branches of the map ${\mathfrak B}_r$ corresponding
to some arbitrary opposite partitions ${\bf w}_-, {\bf w}_+$.
Then, for any initial $ (X,P) \in T^*V_{n,r} $,
$$
{\mathfrak B}_r^- \circ {\mathfrak B}_r^+ (X,P) = (-X,-P),
$$
i.e., double iterations of ${\cal B}_r$ with opposite partitions
multiply $(X, P)$ by $-1$.
\end{thm}

The proof is given in Appendix.

\paragraph{Real trajectories.}
In view of \eqref{moms_r}, for the map ${\mathfrak B}_r$ to be
real-valued it is necessary that $\la_* > a_1,\dots, a_n$, and
even in this case not for any real pair $(X,P)\in T^*V_{(n,r)}$
the matrix equation (\ref{quad_gamma}) has a real solution
$\Gamma$: as follows from the
discrete Legendre transformation \eqref{DLegendreT},
 the real momentum $P$ must belong to the compact set
\begin{equation}\label{compact-set}
\mathbf K(V_{n,r}\times V_{n,r})=
\left\{ A^{1/2}(\la_*)\, \tilde X - \frac 12 X\,\left( \tilde X^T
A^{1/2}(\la_*)X + X^T A^{1/2}(\la_*)\tilde X \right)\right\}.
\end{equation}

\begin{prop}\label{Real}
Assume $\lambda_* > a_1,\dots, a_n$, and let the momentum $P_0$ of the initial real point $(X_0,P_0)\in T^*V_{n,r}$
belong to the set \eqref{compact-set}. Let $L(\lambda_*)$ be evaluated at $(X_0,P_0)$.
Choose the branch of the map $\mathcal B_r$ corresponding to a partition
$$
\mathbf w=\{w_1,\dots,w_r \mid -w_1,\dots,-w_r\},
$$
such that $w_1,\dots,w_r$ are different, $w_i+w_j\ne 0$, $w_i+\bar
w_j\ne 0$, $i,j=1,\dots,r$. Then the trajectory of the discrete
Neumann system \eqref{moms_r*} obtained from $(X_0,P_0)$ by iterating
the above branch is real.
\end{prop}

\noindent{\it Proof.}   Since $w_1,\dots,w_r$ are different and
$w_i+w_j\ne 0$, the corresponding eigenmatrix $\Psi=(\psi_1 \cdots
\psi_r)$ is non-special. Then, according to Proposition
\ref{quad_sol}, the solution $\Gamma=\Upsilon \, \Xi^{-1}$ of
equation \eqref{quad_gamma} with real matrix coefficients
\eqref{blocks} corresponding to $(X_0,P_0)$ is symmetric. On the
other hand, due to the conditions $w_i+\bar w_j\ne 0$, after
setting $\mathfrak U=\mathbf U$, $\mathfrak V=\mathbf V$,
$\mathfrak W=\mathbf W$, the equation \eqref{pot-eq} has the
corresponding hermitian solution $\Gamma$ obtained by the same
eigenmatrix $\Psi$ (see Remark \ref{POTER}). Since the matrix
$\Gamma$ is both symmetric and hermitian, it is a real symmetric
matrix.\hfill $\square$


\begin{rem}{\rm For the case $r=1$, the eigenvalues
$w, -w$ of $L(\lambda^*)$ can be pure real or imaginary. The above
condition gives the real trajectory for real $w$, as in Remark
\ref{REAL}. For the case $r>1$ and real $\lambda^*$, generically
the matrix $L(\lambda^*)$ has a set of four different eigenvalues
invariant with respect to the inversion $w \mapsto -w$ and complex
conjugation $w\mapsto \bar w$. For example, consider the case
$r=2$ and eigenvalues $w,-w,\bar w,-\bar w$. The non-special
eigenmatrix $\Psi$ associated to the partition $ \mathbf
w=\{w_1=w,w_2=\bar w \mid -w_1,-w_2\}$ defines the real branch of
$\mathfrak B_2$. }\end{rem}

Note that with the same initial condition $(X_0,P_0)$ as one
described in Proposition \ref{Real}, we can have a complex
solution of \eqref{quad_gamma}. That is why, it is natural to
consider all objects complexified. In the next subsection we shall
give an algebraic geometrical description of the mapping
$\mathfrak B_r$, which generalize Proposition \ref{shift0}.

\subsection{Algebraic geometric description of ${\mathfrak B}_r$}

Now we describe the correspondence (\ref{moms_r*})
under the eigenvector map
$$
{\cal M}\colon {\cal I}_{red}={\cal I}/SO(r,{\mathbb C})/{\mathbb Z}_2^n \longrightarrow
{\Prym} ( {\mathcal S}', \sigma) \subset {\Jac} ({\mathcal S}')
$$
introduced in \eqref{EM}. As above, for a fixed arbitrary
$\la_*\ne a_i$ choose a partition of eigenvalues
$\mathbf w=\{w_1,\dots,w_r \mid -w_1,\dots,-w_r\}$ of $L(\la_*)$.

Note that the eigenvalues $w_i$ of $L(\la_*)$ and $\hat w_i$ of $\mathbf L(\lambda_*)=a(\lambda_*)L(\lambda_*)$
are related as $\hat w_i=a(\la_*) w_i$, whereas the corresponding eigenvectors can be chosen the same.
Then let $Q_1=(\la_*, a(\la_*) w_1), \dots, Q_r=(\la_*,a(\la_*) w_r)$
be the corresponding points over $\la=\la_*$ on the regularized spectral curve ${\mathcal S}'$ and
$$
\Psi = \begin{pmatrix} \Xi \\ \Upsilon \end{pmatrix} = (\psi(Q_1)
\cdots \psi(Q_r))
$$
be the corresponding non-special eigenmatrix. Let also $\Gamma=
\Upsilon\, \Xi^{-1}$ be the corresponding solution of the matrix
quadratic equation \eqref{quad_gamma}, which fixed the branch of
the map ${\mathfrak B}_r$.

\begin{thm} \label{shift} Under the eigenvector map $\cal M$, the above branch of ${\cal B}_r$ is the
translation on ${\Prym} ( {\mathcal S}', \sigma)$ by the vector given by the degree zero divisor
\begin{equation} \label{shifts}
\mathcal T =\infty_1+ \cdots + \infty_r  -Q_1 - \cdots - Q_r  \,.
\end{equation}
\end{thm}

The proof can be found in Appendix. Note that
$$
\mathcal T+\sigma \mathcal T =  2(\infty_1+ \cdots + \infty_r) -Q_1 - \cdots - Q_r - \sigma Q_1 -
\cdots - \sigma Q_r
$$
which is the divisor of the meromorphic function $1/(\la-\la_*)$
on ${\mathcal S}'$. Hence $\mathcal T+\sigma \mathcal T\equiv 0$,
and $\mathcal T$ indeed belongs to ${\Prym} ( {\mathcal S}',
\sigma)$.

We also note that the expression \eqref{shifts} is a direct
generalization of the one-point translation described in
Proposition \ref{shift0} for the case $r=1$. In the continuous
limit $\la_*\to\infty$ we have $Q_1+\dots+Q_r\longmapsto
\infty_1+\dots+\infty_r$, hence the shift vector $T={\cal
A}(\mathcal T)$ tends to zero.

\paragraph{Growth of the map.}
Theorem \ref{shift} says that the shift $\mathcal T$ on ${\Prym} (
{\mathcal S}', \sigma)$ does not depend on the step of iteration
of ${\mathfrak B}_r$, but only on the choice of partition ${\bf
w}=\{w_1,\dots,w_r \mid -w_1,\dots,-w_r\}$. The opposite
partitions $ {\bf w}_-, {\bf w}_+$ produce opposite shifts
$\mathcal T_-, \mathcal T_+$ such that $\mathcal T_- + \mathcal
T_+ \equiv 0$. Hence the $2^r$ branches of the map ${\mathfrak
B}_r $ and of the composition ${\mathcal  M} \circ {\mathfrak
B}_r$ can conditionally be divided into $2^{r-1}$ ''forward'' and
$2^{r-1}$ ''backward'' branches. Clearly, double iterations of $
{\cal M} \circ  {\mathfrak B}_r$ with opposite partitions give the
original value of ${\cal M} (X,P)$. This property also holds for
the map ${\mathfrak B}_r$ itself, but not completely: due to
Theorem \ref{reiteration}, the corresponding double iterations of
${\mathfrak B}_r$ give $(-X,-P)$ and not $(X,P)$.

In ${\mathbb C}^l$, the universal covering of ${\Prym} ( {\mathcal
S}', \sigma)$, all the iterations of ${\cal M} \circ {\mathfrak
B}_r$ form a lattice $\Xi$ of rank $\le 2^{r-1}$. Due to the
definition of the vectors ${\cal A} (\mathcal T)$, for $r>2$ not
all of them are linearly independent, hence the lattice fits into
a linear subspace of dimension less than $2^{r-1}$. For example,
let $r=3$ and denote $\infty=\infty_1+\infty_2 +\infty_3$. The
lattice $\Xi$ is generated by 4 shift vectors ${\cal A} ({\mathcal
T}_1),\dots, {\cal A} ({\mathcal T}_4)$ with
\begin{eqnarray*}
&& {\mathcal T}_1 =-Q_1-Q_2-Q_3+\infty, \qquad\qquad
{\mathcal T}_2 =-\sigma (Q_1)-Q_2-Q_3+\infty, \\
&& {\mathcal T}_3 =-Q_1-\sigma(Q_2)-Q_3+\infty, \quad \qquad
{\mathcal T}_4 =-Q_1 -Q_2- \sigma(Q_3)+ \infty,
\end{eqnarray*}
and one can see that ${\cal A} (\mathcal T_1)+{\cal A} (\mathcal
T_2)+{\cal A} (\mathcal T_3) + {\cal A}({\mathcal T}_4)=0$.

The above implies that iterations of the ''forward'' branches of ${\cal M} \circ
{\mathfrak B}_r$ give a linear growth of the images of the map.

\section{Conclusion}

We have seen that the Neumann systems on $V_{n,r}$ (continuous and discrete) inherit or naturally generalize
the basic properties of the classical
Neumann system on $S^{n-1}$ and, therefore, of the (odd) Jacobi--Mumford systems: the structure of the Lax matrices
(symplectic), the spectral curve (with involution), the equations of motion, linearization on Abelian varieties,
and, in the discrete case, the formula for the translation $\cal T$ on them.

For this reason, we believe that the Hamiltonian systems on $T^*V_{n,r}$ we consider
represent one of the most natural matrix generalizations of the odd Jacobi--Mumford systems.

On the other hand, such generalizations are quite specific since, by construction, the corresponding matrix residua
${\cal N}_i$ in \eqref{Nr*} have rank 1.  So, it is also natural to consider extensions of
the Neumann systems whose Lax matrices $L(\la)$ remain to be symplectic, but with residua of a higher rank at $\la=a_i$.

To reach a full generality, one can introduce the space ${\cal E}$ of three $r \times r$ matrix polynomials
\begin{align*}
{\bf U} (\la) &= \la^n {\bf I}_r + U_1 \la^{n-1}+\cdots+ U_n,  \\
{\bf V} (\la)&= V_0 \la^n + V_1 \la^{g-1}+ \cdots+V_{n}, \\
{\bf W} (\la) &=\la^{n+1}{\bf I}_r + W_0 \la^{n} + W_1 \la^{n-1} + \cdots+ W_n ,
\end{align*}
where $U_i, W_j$ are symmetric and $V_0$ is {\it skew-symmetric} arbitrary coefficients, and consider a hierarchy of flows
on $\cal E$ given by the Lax pairs
$$
 \frac{d}{dt}{\mathbb L}(\la) = [{\mathbb L}(\la), {\mathbb N}(\la) ], \quad
  {\mathbb L}(\la) = \begin{pmatrix} {\bf V}(\la) & {\bf U}(\la) \\
                 {\bf W} (\la) & - {\bf V}^T  (\la)  \end{pmatrix}.
$$
In particular, ${\mathbb N}(\la)$ can be choosen in the form similar to \eqref{LA4}
$$
   {\mathbb N}(\la) = \begin{pmatrix} V_0 & {\bf I}_r \\
                 \la {\bf I}_r + \Lambda & - V_0   \end{pmatrix}, \qquad \Lambda= W_0-U_1 ,
$$
which implies that $V_0$ is a matrix first integral of the flow. One can prove that in the case $V_0=0$ the matrix
$\Lambda$ satisfies a stationary reduction of one of the equations of the $r\times r$ symmetric matrix KdV hierarchy
considered, in particular, in \cite{AF}.

The spectral curve ${\cal S}=\{|{\mathbb L}(\lambda)-w{\bf I}_{2r}|=0\}$ has the involution
$\sigma\, : (\la,w)\to (\la, -w)$ and, as above, its complete regularization ${\mathcal S}'$ has $r$ infinite points.
It is then natural to conjecture that
generic complex invariant manifolds of the flows on $\cal E$ are open subsets of non-compact extensions
of the Prym varieties ${\Prym} ( {\mathcal S}', \sigma)\subset \Jac({\mathcal S}')$, and
if $V_0=0$, such manifolds are the Prym varieties themselves.

Similarly, one can introduce a family of multi-valued maps
$$
{\bf B}_{\la_*} \, :\, {\cal E}\to {\cal E}, \quad (U_i, V_j, W_k) \to (\tilde U_i, \tilde V_j, \tilde W_k), \quad
\la_*\in {\mathbb C}
$$
defined by the intertwining relation
\begin{gather}
\tilde{\mathbb L}(\la) M(\la,\la_*) =M(\la,\la_*) {\mathbb L} (\la) , \label{mat_int} \\
{\mathbb L}(\la) = \begin{pmatrix} {\bf V}(\la) & {\bf U}(\la) \\
                 {\bf W} (\la) & - {\bf V}^T  (\la)  \end{pmatrix}, \quad
M (\la, \la_*) = \begin{pmatrix} - \Gamma & {\bf I}_r \\
 (\la - \la_*) {\bf I}_r +  \Gamma^2  &  - \Gamma  \end{pmatrix} , \notag
\end{gather}
where $\tilde {\mathbb L} (\la)$ depends on the polynomials
$ \tilde {\bf U} (\la), \tilde {\bf V} (\la), \tilde {\bf W} (\la)$ in the same way as
 ${\mathbb L} (\la)$ depends on ${\bf U} (\la),\dots, {\bf W} (\la)$, and
$\Gamma$ is an $r \times r$ symmetric matrix determined from the compatibility of the left- and right
hand sides of \eqref{mat_int}. This condition leads to the following $r \times r$ matrix quadratic equation
\begin{equation} \label{MQE}
\Gamma {\bf U}(\la_*) \Gamma + \Gamma {\bf V}(\la_*)+ {\bf V}^T(\la_*) \Gamma - {\bf W} (\la_*)  =0 .
\end{equation}
Once one of the $2^r$ solutions $\Gamma$ of \eqref{MQE} is fixed by using Proposition \ref{quad_sol}, the map
${\bf B}_{\la_*}$ is defined uniquely. Like in the continuous case, one finds that $\tilde V_0=- V^T_0$,
which implies preservation of the skew-symmetric matrix $V_0$ under any branch of ${\bf B}_{\la_*}$.
It is expected that the branches of ${\bf B}_{\la_*}$ are given by translations on $\Jac({\mathcal S}')$
described by Theorem \ref{shift}.

\subsection*{Acknowledgments} The authors are grateful to M. Alberich for making independently
a series of hard calculations of the singularities of the spectral curve ${\cal S}'$ and of its genus.
We also thank L. Gavrilov and J. C. Naranjo for valuable remarks.

Y.F acknowledges support of the Spanish MINECO-FEDER Grants
MTM2012-31714,  MTM2012-37070. The research of B. J. was supported
by the Serbian Ministry of Science, Project 174020, Geometry and
Topology of Manifolds, Classical Mechanics and Integrable
Dynamical Systems.


\section{Appendix: Technical proofs}

\noindent{\it Proof of Lemma} \ref{Lem1}. 1) Take two distinct
eigenvalues $w, w'$ of ${\bf L}(\la)$ and the corresponding
eigenvectors $\psi(\la,w), \psi^*(\la,w') $. Then, on the one
hand,
$$
 \left( \psi^*(\la,w') \right)^T {\bf L} (\la) \psi(\la,w) =  w \left( \psi^*(\la,w') \right)^T \psi(\la,w).
$$
On the other hand, in view of \eqref{eigen_transpose}, the same
product equals
$$
\left( \psi^*(\la,w') \right)^T {\bf L}(\la) \psi(\la,w) =
\left({\bf L}^T (\la) \psi^*(\la,w') \right)^T \psi(\la,w)  = w'
\left( \psi^*(\la,w') \right)^T \psi(\la,w).
$$
Since $w' \ne w$, the difference of the above two equations yields
$\left( \psi^*(\la,w') \right)^T \psi(\la,w)=0$ for any
$(\la,w)\ne (\la,w')$ and, by continuity, $\left( \psi^* (\la,w)
\right)^T \psi(\la,w)=0$ when $(\la,w)\in {\cal B}$. Hence, $F(P)$
vanishes at all the branch points $\cal B$.

Next, since the components $\chi (P), \xi (P)$ of $\psi(P)$ have
poles only at ${\cal D}$ and, by their definition, the components
of $\psi^*(P)$ have poles only at $\sigma {\cal D}$, the function
$F(P)$ has poles only at ${\cal D}+\sigma {\cal D}$. Note that the
latter points are all finite, that is, distinct from
$\infty_1,\dots, \infty_r\in {\cal B}$, if we impose an
appropriate normalization, for example $\psi^1 (P)+\cdots +
\psi^{2r}(P)=1$.

Finally, note that, apart from simple zeros at $\cal B$, the
function $F(P)$ cannot have other zeros on ${\mathcal S}'$,
because, due to the Riemann--Hurwitz formula, the genus of the
curve $S'$, as a $2r$-fold covering of $\mathbb P$, equals
$$
g=\mathrm{gen}(\mathcal S')=2r(\mathrm{gen}(\mathbb P)-1)+ \mathrm{deg}({\cal B})/2+1=-2r+ \mathrm{deg}({\cal B})/2+1,
$$
then we have deg$({\cal
B})=2(g+2r-1)$, and, in view of \eqref{deg},
$$
\text{deg} ({\cal D} + \sigma {\cal D}) =2 \text{deg} ({\cal D})=
2(g+2r-1) =\text{deg} ({\cal B}) .
$$

2) Since $\psi(\sigma P) = \bar\psi(P) = (\bar\chi, \bar\xi)^T$
and $\psi^*(\sigma P)$ satisfies $L^T (\la) \psi^*(\sigma P) = - w
\psi^*(\sigma P)$, we have $\psi^*(\sigma P) = (-\xi, \chi)^T$. As
a result,
$$
 F(\sigma P) = (\psi^*(\sigma P))^T \psi(\sigma P)= -\bar\chi \xi + \chi\bar\xi = - F(P),
$$
which completes the proof of the lemma. \hfill$\square$

\medskip

\noindent{\it Proof of Theorem \ref{linear2}.} At the first step,
developing the idea of proof of Theorem 6.39 in \cite{AMV}, we
show that for any differential $\varOmega$ from the generalized
Abel map \eqref{GAM},
\begin{equation}\label{amv3}
\frac{d}{dt}\bigg |_{t=0}\int_{\mathcal D_0}^{\mathcal
D_t}\varOmega= -\lim_{t\to 0} \sum_{s=1}^{r}
\frac{d}{dt}\int_{Q_s}^{P_s(t)}\varOmega \, ,
\end{equation}
where $P_1(t),\dots, P_r(t) \in {\cal S}'$ are the solutions of
the equation $f(P,0)={1}/{t}$ for small $t$ and $Q_1,\dots,Q_r$
are any finite fixed points.

Namely, let $(U_0,U_\infty)$ denote an open cover of $\mathcal S'$
such that $U_\infty\setminus U_0$ is a neighborhood of infinite
points $\infty_1,\dots,\infty_r$, while $U_0\setminus U_\infty$ is
a neighborhood of the support of $\mathcal D_0$. Apart from the
divisor $\mathcal D_t$, for a small $t\ne 0$ consider the divisor
$\mathcal D'_t$ defined by the relation
$$
(1-t \, f(P,0))_{U_0}=\mathcal D'_t-\mathcal D_0,
$$
where $(1-tf(P,0))_{U_0}$ is a part of the divisor $(1-tf(P,0))$
with a support belonging to $U_0$. If $P_1(t),\dots,P_r(t)\in
U_\infty\setminus U_0$ are the solutions of $f(P,0)={1}/{t}$ such
that $\lim_{t\to 0} P_s(t) = \infty_s$, then
$$
(1-t f(P,0))=\mathcal D'_t-\mathcal D_0+
P_1(t)+\dots+P_r(t)-(\infty_1+\dots+\infty_r).
$$

Next, introduce the function $g(P,t)=(1-t f(P,0))^2/\lambda$
satisfying  $g(\infty_s^+)=g(\infty_s^-)=t^2\ne 0,\infty$,
$s=1,\dots,[r/2]$. Then, the divisor of $g$ gives zero in
$\widetilde\Jac({\cal S}', \infty)$:
\begin{align*}
(g) & =2(1-t f(P,0))-(\lambda)\\
&= 2(1-t f(P,0)) +2(\infty_1+\dots+\infty_r)- \mathcal O \\
&= 2\mathcal D'_t-2\mathcal D_0+2(P_1(t)+\dots+P_r(t))-\mathcal O
\equiv_\infty 0 \, ,
\end{align*}
where, as above, the divisor $\mathcal O= O_1+\dots+O_{2r}$ is the
preimage of $\la=0$ on ${\cal S}'$. That is, for any $t$,
$\tilde{\mathcal A}\left(2\mathcal D'_t-2\mathcal
D_0+2(P_1+\dots+P_r)-\mathcal O\right)\in\widetilde\varLambda$
and, therefore,
$$
2\lim_{t\to 0}\frac{d}{dt}\int_{\mathcal D_0}^{\mathcal
D'_t}\varOmega=-\sum_{s=1}^{r} \lim_{t\to
0}\frac{d}{dt}\Big(\int_{O_s}^{P_s(t)}\varOmega+\int_{O_{2s}}^{P_s(t)}\varOmega\Big),
$$
where $\varOmega$ is a holomorphic or a meromorphic differential
of 3rd kind with the poles at the infinite points.

Since $\mathcal O$ does not depend on $t$, the above relation
implies
\begin{equation}\label{amv1}
\lim_{t\to 0}\frac{d}{dt}\int_{\mathcal D_0}^{\mathcal
D'_t}\varOmega=-\sum_{s=1}^{r} \lim_{t\to
0}\frac{d}{dt}\int_{Q_s}^{P_s(t)}\varOmega,
\end{equation}
with any fixed points $Q_1,\dots,Q_r\in U_\infty\setminus U_0$
close to $\infty_1,\dots,\infty_r$, respectively.

Further, as in Proposition 6.37 of \cite{AMV}, for small t we have
\begin{equation}\label{amv2}
\int_{\mathcal D_t}^{\mathcal D'_t} \varOmega=O(t^2) .
\end{equation}
Indeed, all the arguments of that proposition can be applied in
our situation. Although in \cite{AMV} the property \eqref{amv2}
was proved for holomorphic differentials, the proof uses only
their restriction to $U_0$, so it is applicable for meromorphic
differentials $\varOmega_k$ as well.

As a result, from \eqref{amv1}, \eqref{amv2} we get the relation
\eqref{amv3}.

At the second step we use the following property.

\begin{lem}\label{RES}
If $\varOmega$ is a holomorphic near $\infty_s$, then
\begin{equation*}
\lim_{t\to 0}\frac{d}{dt}\int_{Q_s}^{ P_s(t)
}\varOmega=\mathrm{Res}_{\infty_s} f\varOmega .
\end{equation*}
On the other hand, if $\varOmega=\varOmega_k$ is a normalized
anti--invariant meromorphic differential with the residia $\pm 1$
at $\infty_k^+, \infty_k^-$, then
\begin{equation*}
\lim_{t\to
0}\frac{d}{dt}\Big(\int_{Q_k}^{P_k(t)}\varOmega_k+\int_{Q_{k+[r/2]}}^{P_{k+[r/2]}(t)}\varOmega_k\Big)=\mathrm{Res}_{\infty_k^+}
f\varOmega_k+\mathrm{Res}_{\infty_k^-} f\varOmega_k+2\kappa\nu_k.
\end{equation*}
for any $k=1,\dots,[r/2]$.
\end{lem}

Since the residia $\mathrm{Res}_{\infty_s} f\varOmega$ do not
depend on $t$, applying the above Lemma to \eqref{amv3} with
$\varOmega =
\omega_1,\dots,\omega_g,\varOmega_1,\dots,\varOmega_{[r/2]}$, we
prove the relation \eqref{new-equation} for the case of
anti-invariant meromorphic differentials $\varOmega_k$. However, a
general meromorphic differential with simple poles at $\infty_k^+,
\infty_k^-$ is a linear combination of $\varOmega_k$ and
$\omega_1,\dots,\omega_g$, hence Theorem \ref{linear2} holds for
the general case as well. \hfill $\Box$

\medskip

\noindent{\it Proof of Lemma} \ref{RES}. The solutions
$P_1(t),\dots,P_r(t)$ of the equation $f(P,0)=1/t$ correspond to
the following expansions of their local coordinates $\tau$
\begin{eqnarray}
\nonumber P_k(t)\, :\; && \tau=t+\kappa \nu_k \, t^2 +O(t^3),  \\
P_{[r/2]+k}(t)\, :\;&& \label{exp-p} \tau =t -\kappa \nu_k \, t^2 + O(t^3), \qquad k=1,\dots,[r/2],\\
\nonumber P_r(t)\,\, :\;&&  \tau=t + O(t^3) \qquad
\qquad\qquad\qquad (\text{if {\it r} is odd}).
\end{eqnarray}
This can be proved by substituting the expansions \eqref{exp-f} to
$f(P,0)=1/t$ and using the implicit function theorem.

Then, for a holomorphic differential $\varOmega$ with the
expansion $\big(\varphi_{s}+O(\tau)\big)d\tau$ near $\infty_s$, we have
\begin{equation*}
\lim_{t\to 0}\frac{d}{dt}\int_{Q_s}^{P_s(t)}\varOmega = \lim_{t\to
0}\big(\varphi_{s}+O(\tau(t))\big)\dot\tau
(t)=\varphi_{s}=\mathrm{Res}_{\infty_s} f\varOmega.
\end{equation*}
For a meromorphic differential $\varOmega_k$ with the behavior
\eqref{exp_Om} near $\infty_k^+$ and $\infty_k^-$, the expansions
\eqref{exp-p} imply $\dot\tau (t) = 1\pm  \kappa \nu_k\, 2t +
O(t^2)$ and
\begin{align*}
& \lim_{t\to 0}\frac{d}{dt}\Big(\int_{Q_k}^{P_k(t)}\varOmega_k+\int_{Q_{k+[r/2]}}^{P_{k+[r/2]}(t)}\varOmega_k\Big)\\
&\qquad=\lim_{t\to 0}\Big[\Big(\frac{1}{\tau (t)} + \phi_{k,k} +
O(\tau) \Big) \dot\tau(t) +\Big(- \frac {1}{\tau (t)} +
\phi_{k,k} +
O(\tau) \Big) \dot\tau (t) \Big]\\
&\qquad =2\phi_{k,k}+4\kappa\nu_k=\mathrm{Res}_{\infty_k^+}
f\varOmega_k+\mathrm{Res}_{\infty_k^-} f\varOmega_k+2\kappa\nu_k.
\end{align*}
\hfill$\Box$

\medskip

\noindent{\it Proof of Theorem} \ref{reiteration}. Expressing $X$
from the second matrix equation in (\ref{moms_r}) and using the
condition $X^T X = {\bf I}_r$, we obtain the following alternative
equation for the matrix multiplier $\Gamma$,
\begin{gather} \label{quad_g2}
 \Gamma  \tilde {\bf U} \Gamma - \Gamma \tilde{\bf V} - \tilde{\bf V}^T \Gamma - \tilde{\bf W}=0, \\
\tilde {\bf U}= \tilde X^T A^{-1}(\la_*) \tilde X , \quad \tilde
{\bf V}= \tilde X^T A^{-1}(\la_*) \tilde P, \quad \tilde {\bf W}=
{\bf I}_r - \tilde P^T A^{-1}(\la_*)\tilde P  \nonumber
\end{gather}

According to Proposition \ref{quad_sol}, all symmetric matrix
solutions of (\ref{quad_g2}) have the form $\Gamma= {\hat\Upsilon
} \, {\hat\Xi}^{-1}$, where ${\hat\Xi}, {\hat\Upsilon }$ are
halves of a nonspecial eigenmatrix $\hat\Psi$ of $\widehat L$,
$$
\hat \Psi = \begin{pmatrix}
\hat\Xi \\ \hat\Upsilon  \end{pmatrix}, \qquad
\widehat L= \begin{pmatrix} -\tilde {\bf V} & \tilde {\bf U} \\
\tilde {\bf W} & \tilde {\bf V}^T \end{pmatrix}.
$$

To see how the solutions of (\ref{quad_gamma}) and \eqref{quad_g2}
are related, let us fix a solution $\Gamma^*$ of the original
equation (\ref{quad_gamma}) corresponding to a certain partition
${\bf w}^* =\{ w\mid -w\}$. Let $\tilde X^*, \tilde P^*$ be the
corresponding image, $\tilde {\bf U}^*, \tilde {\bf V}^*, \tilde
{\bf W}^*$ be the corresponding coefficients in the equation
\eqref{quad_g2}, and $\{\hat\Gamma \}$ be the set of its
solutions. Then $\Gamma^* \in \{\hat\Gamma \}$.

On the other hand, let $ \{\tilde\Gamma\}$ be the complete set of
solution of the original matrix equation (\ref{quad_gamma}) but
with the coefficients defined by the {\it tilded} variables
$\tilde X^*,\tilde P^*$. It is seen that if $\tilde\Psi$ is the
nonspecial eigenmatrix of $\tilde L(\la_*)$,
$$
\tilde\Psi =
\begin{pmatrix} \tilde\Xi \\ \tilde\Upsilon  \end{pmatrix}, \qquad
\tilde L(\la_*)=
\begin{pmatrix} \tilde {\bf V} & \tilde {\bf U} \\ \tilde {\bf W} & -\tilde {\bf V}^T
\end{pmatrix},
$$
corresponding to a partition $\{w_1,\dots,w_r \mid-w_1,\dots,-w_r\}$, then $(-\tilde\Xi^T \tilde\Upsilon^T)^T$
is the nonspecial eigenmatrix of
the above matrix $\widehat L$ corresponding to the opposite
partition $\{-w_1,\dots,-w_r \mid w_1,\dots, w_r\}$. Indeed,
$$
 \begin{pmatrix} \tilde {\bf V} & \tilde {\bf U} \\ \tilde {\bf W} & -\tilde {\bf V}^T \end{pmatrix}
 \begin{pmatrix} \tilde\Xi \\ \tilde\Upsilon  \end{pmatrix} = w_s \begin{pmatrix} \tilde\Xi \\ \tilde\Upsilon  \end{pmatrix}
\,\,\,\text{implies}\,\,\,
\begin{pmatrix} -\tilde {\bf V} & \tilde {\bf U} \\ \tilde {\bf W} & \tilde {\bf V}^T \end{pmatrix}
\begin{pmatrix}
-\tilde\Xi \\ \,\,\tilde\Upsilon
\end{pmatrix}= -w_s
\begin{pmatrix}
-\tilde\Xi \\ \,\,\tilde\Upsilon
\end{pmatrix},
$$
for $s=1,\dots,r$. Hence the set $\{\tilde\Gamma\}$ contains the
solution $-\Gamma^*$.

To complete the proof, consider the tilded version of the
equations (\ref{moms_r}),
\begin{equation} \label{tilded_moms}
\tilde P^*  =\;  A^{1/2}(\la_*)\, \tilde{\tilde X} - \tilde X^*\,
\tilde\Gamma(\la_*), \quad \tilde{\tilde P} = - A^{1/2}(\la_*)\,
\tilde X^* + \tilde{\tilde X}\, \tilde\Gamma(\la_*) .
\end{equation}
Setting here $\tilde \Gamma= - \Gamma^*$ and assuming that
$\tilde{\tilde P}=-P$, $\tilde{\tilde X}=-X$, we see that the
first (second) equation of (\ref{tilded_moms}) becomes the second
(first) equation in (\ref{moms_r}). Hence, the equations
(\ref{tilded_moms}) for $\tilde{\tilde X}, \tilde{\tilde P}$ have
the solution $(-X,-P)$.\hfill $\square$

\medskip

\noindent{\it Proof of Theorem} \ref{shift}. As follows from the
intertwining relation \eqref{disr_Lax_N}, if $\psi(P)$ is an
eigenvector of $L(\la)$, then
$$
 \widehat \psi (P) = M(\la, \la_*)\psi(P) = \begin{pmatrix} - \Gamma & {\bf I}_r \\
 (\la - \la_*) {\bf I}_r +  (\Gamma)^2 &  - \Gamma \end{pmatrix}  \psi(P)
$$
is an eigenvector of $\tilde L(\la)$ with the same eigenvalue.
Note that whereas $\psi(P)$ is normalized, the above $\widehat
\psi(P)$ is not, so we consider normalized eignevector
$\tilde\psi(P)=f^{-1}(P) \widehat\psi(P)$, $f=\langle \alpha,
\widehat\psi(P)\rangle$ for a normalization $\alpha\in{\mathbb
P}^{2r-1}$.

We compare the divisors $\cal D, \tilde{\cal D}$ of poles of $\psi
(P), \tilde \psi (P)$ for the chosen $\Gamma$. First, note that
$\det M(\la, \la_*)=(\la-\la_*)^r$. Hence, apart from the points
of ${\cal S}'$ over $\la=\la_*$ and $\la=\infty$, $M(\la, \la_*)$
is non-degenerate. Therefore, apart from these points, the
divisors of poles of $\psi(P)$ and $\widehat\psi(P)$ coincide.

Without loss of generality, assume that $\psi(P)$ is normalized in
the same way as in Proposition 4.1. Then in the neighborhood of
the infinite points $\infty_1,\dots, \infty_r$ these components
have the expansions \eqref{exp_psi} and, near each $\infty_s$ with
the local coordinate $\tau=1/\sqrt{\la}$, the following expansion
holds
\begin{gather*}
 \widehat \psi (P) = \begin{pmatrix} - \Gamma & {\bf I}_r \\
 \tau^{-2} {\bf I}_r +  (\Gamma)^2 + O(1) {\bf I}_r &  - \Gamma \end{pmatrix}
\begin{pmatrix} {\bf v}_s \tau + O(\tau^2) \\ {\bf v}_s + O(\tau ) \end{pmatrix}
 = \begin{pmatrix} {\bf v}_s + O(\tau ) \\ \tau^{-1} {\bf v}_s - \Gamma {\bf v}_s + O(\tau)
 \end{pmatrix},
\end{gather*}
where, as in \eqref{expansion1}, $\mathbf v_{j+[r/2]}=\bar{\mathbf
v}_j$, $j=1,\dots,[r/2]$, and, in the case when $r$ is odd,
$\mathbf v_r=\mathbf v_0$.

Therefore, in contrast to $\psi (P)$, some components of
$\widehat\psi (P)$ have a first order pole at $\infty_s$. Next,
observe that the eigenvectors $\psi(Q_1), \dots, \psi(Q_r)$ that
form the eigenmatrix $\Psi$, span the kernel of $M(\la_*, \la_*)$.
Indeed, in view of the relation $\Gamma= \Upsilon \Xi^{-1}$,
$$
M(\la_*, \la_*) \Psi =
\begin{pmatrix} - \Gamma & {\bf I}_r \\
 \Gamma^2 &  - \Gamma \end{pmatrix} \begin{pmatrix} \Xi  \\  \Upsilon  \end{pmatrix} =
\begin{pmatrix} - \Gamma & {\bf I}_r \\
 \Gamma^2 &  - \Gamma \end{pmatrix} \begin{pmatrix} {\bf I}_r  \\  \Gamma  \end{pmatrix}\Upsilon = \bf{0}.
$$

The products of $M(\la_*, \la_*)$ with the other eigenvectors
$\psi(\sigma Q_1)$, $\dots$, $\psi(\sigma Q_r)$ are non-zero. It
follows that the divisor of the above normalizing factor $f(P)$ is
$$
 (f) = Q_1+\cdots +Q_r + {\cal R} -{\cal D} -\infty_1-\cdots - \infty_r
$$
for a certain effective divisor ${\cal R}$. Then the divisor of
poles $\tilde{\mathcal D}$ of $\tilde\psi(P)=\widehat\psi(P)/f(P)$
equals ${\cal R}$. Indeed, the zeros of $\widehat\psi(P)$ and
$f(P)$ at $Q_1+\dots+Q_r$ as well as their poles at $\mathcal
D+\infty_1+\dots+\infty_r$ cancel each other. Since $f(P)$ is
meromorphic on ${\cal S}'$, we conclude that $\tilde{\cal D}$ is
equivalent to
$$
{\cal D} + \infty_1+\cdots+ \infty_r - Q_1 -\cdots - Q_r ,
$$
which implies that the images of $\cal D$, $\tilde{\cal D}$ in
${\Jac} ({\mathcal S}')$ differ by the translation $\mathcal T$ in
\eqref{shifts}.
 \hfill$\square$

Yuri N. Fedorov

Department of Mathematics

Polytechnic university of Catalonia

Barcelona, E-08028 Spain

e-mail: Yuri.Fedorov@upc.edu

\

Bo\v zidar Jovanovi\' c

Mathematical Institute SANU

Serbian Academy of Science and Arts

Kneza Mihaila 36, 11000, Belgrad, Serbia

e-mail: bozaj@mi.sanu.ac.rs
\end{document}